\documentclass[12pt]{spieman}  

\usepackage{amsmath,amsfonts,amssymb}
\usepackage{graphicx}
\usepackage{setspace}
\usepackage{tocloft}
\usepackage{longtable}

\definecolor{red}{rgb}{1.0,0.0,0.0}

\title{Revised Astrometric Calibration of the Gemini Planet Imager}
\author[a,b]{Robert J. De Rosa}
\author[c]{Meiji M. Nguyen}
\author[d]{Jeffrey Chilcote}
\author[b]{Bruce Macintosh}
\author[e]{Marshall D. Perrin}
\author[f]{Quinn Konopacky}
\author[g]{Jason J. Wang}
\author[c,h]{Gaspard Duch\^ene}
\author[b]{Eric L. Nielsen}
\author[h,i]{Julien Rameau}
\author[j]{S. Mark Ammons}
\author[k]{Vanessa P. Bailey}
\author[l]{Travis Barman}
\author[m,n]{Joanna Bulger}
\author[o]{Tara Cotten}
\author[i]{Rene Doyon}
\author[c]{Thomas M. Esposito}
\author[p]{Michael P. Fitzgerald}
\author[q]{Katherine B. Follette}
\author[r,s]{Benjamin L. Gerard}
\author[t]{Stephen J. Goodsell}
\author[c]{James R. Graham}
\author[u]{Alexandra Z. Greenbaum}
\author[a]{Pascale Hibon}
\author[v]{Li-Wei Hung}
\author[w]{Patrick Ingraham}
\author[c,x]{Paul Kalas}
\author[p]{James E. Larkin}
\author[f]{J\'er\^ome Maire}
\author[x]{Franck Marchis}
\author[y]{Mark S. Marley}
\author[s,r]{Christian Marois}
\author[z,1]{Stanimir Metchev}
\author[k]{Maxwell A. Millar-Blanchaer}
\author[2]{Rebecca Oppenheimer}
\author[j]{David Palmer}
\author[3]{Jennifer Patience}
\author[j]{Lisa Poyneer}
\author[e]{Laurent Pueyo}
\author[e]{Abhijith Rajan}
\author[4]{Fredrik T. Rantakyr\"o}
\author[b]{Jean-Baptiste Ruffio}
\author[5]{Dmitry Savransky}
\author[3]{Adam C. Schneider}
\author[e]{Anand Sivaramakrishnan}
\author[o]{Inseok Song}
\author[e]{Remi Soummer}
\author[w]{Sandrine Thomas}
\author[k]{J. Kent Wallace}
\author[q]{Kimberly Ward-Duong}
\author[6]{Sloane Wiktorowicz}
\author[7]{Schuyler Wolff}
\affil[a]{European Southern Observatory, Alonso de Cordova 3107, Vitacura, Santiago, Chile}
\affil[b]{Kavli Institute for Particle Astrophysics and Cosmology, Stanford University, Stanford, CA 94305, USA}
\affil[c]{Department of Astronomy, University of California, Berkeley, CA 94720, USA}
\affil[d]{Department of Physics, University of Notre Dame, 225 Nieuwland Science Hall, Notre Dame, IN 46556, USA}
\affil[e]{Space Telescope Science Institute, Baltimore, MD 21218, USA}
\affil[f]{Center for Astrophysics and Space Science, University of California San Diego, La Jolla, CA 92093, USA}
\affil[g]{Department of Astronomy, California Institute of Technology, Pasadena, CA 91125, USA}
\affil[h]{Univ. Grenoble Alpes/CNRS, IPAG, F-38000 Grenoble, France}
\affil[i]{Institut de Recherche sur les Exoplan{\`e}tes, D{\'e}partement de Physique, Universit{\'e} de Montr{\'e}al, Montr{\'e}al QC, H3C 3J7, Canada}
\affil[j]{Lawrence Livermore National Laboratory, Livermore, CA 94551, USA}
\affil[k]{Jet Propulsion Laboratory, California Institute of Technology, Pasadena, CA 91109, USA}
\affil[l]{Lunar and Planetary Laboratory, University of Arizona, Tucson, AZ 85721, USA}
\affil[m]{Institute for Astronomy, University of Hawaii, 2680 Woodlawn Drive, Honolulu, HI 96822, USA}
\affil[n]{Subaru Telescope, NAOJ, 650 North A{'o}hoku Place, Hilo, HI 96720, USA}
\affil[o]{Department of Physics and Astronomy, University of Georgia, Athens, GA 30602, USA}
\affil[p]{Department of Physics \& Astronomy, University of California, Los Angeles, CA 90095, USA}
\affil[q]{Physics and Astronomy Department, Amherst College, 21 Merrill Science Drive, Amherst, MA 01002, USA}
\affil[r]{University of Victoria, 3800 Finnerty Road, Victoria, BC, V8P 5C2, Canada}
\affil[s]{National Research Council of Canada Herzberg, 5071 West Saanich Rd, Victoria, BC, V9E 2E7, Canada}
\affil[t]{Gemini Observatory, 670 N. A'ohoku Place, Hilo, HI 96720, USA}
\affil[u]{Department of Astronomy, University of Michigan, Ann Arbor, MI 48109, USA}
\affil[v]{Natural Sounds and Night Skies Division, National Park Service, Fort Collins, CO 80525, USA}
\affil[w]{Large Synoptic Survey Telescope, 950N Cherry Avenue, Tucson, AZ 85719, USA}
\affil[x]{SETI Institute, Carl Sagan Center, 189 Bernardo Avenue,  Mountain View CA 94043, USA}
\affil[y]{NASA Ames Research Center, MS 245-3, Mountain View, CA 94035, USA}
\affil[z]{Department of Physics and Astronomy, Centre for Planetary Science and Exploration, The University of Western Ontario, London, ON N6A 3K7, Canada}
\affil[1]{Department of Physics and Astronomy, Stony Brook University, Stony Brook, NY 11794-3800, USA}
\affil[2]{Department of Astrophysics, American Museum of Natural History, New York, NY 10024, USA}
\affil[3]{School of Earth and Space Exploration, Arizona State University, PO Box 871404, Tempe, AZ 85287, USA}
\affil[4]{Gemini Observatory, Casilla 603, La Serena, Chile}
\affil[5]{Sibley School of Mechanical and Aerospace Engineering, Cornell University, Ithaca, NY 14853, USA}
\affil[6]{The Aerospace Corporation, 2310 E. El Segundo Blvd., El Segundo, CA 90245, USA}
\affil[7]{Leiden Observatory, Leiden University, 2300 RA Leiden, The Netherlands}

\newcommand{\farcs}{\mbox{\ensuremath{.\!\!^{\prime\prime}}}}
\cftpagenumbersoff{figure}
\cftpagenumbersoff{table} 
\begin{document} 
\maketitle

\begin{abstract}
We present a revision to the astrometric calibration of the Gemini Planet Imager (GPI), an instrument designed to achieve the high contrast at small angular separations necessary to image substellar and planetary-mass companions around nearby, young stars. We identified several issues with the GPI Data Reduction Pipeline (DRP) that significantly affected the determination of angle of north in reduced GPI images. As well as introducing a small error in position angle measurements for targets observed at small zenith distances, this error led to a significant error in the previous astrometric calibration that has affected all subsequent astrometric measurements. We present a detailed description of these issues, and how they were corrected. We reduced GPI observations of calibration binaries taken periodically since the instrument was commissioned in 2014 using an updated version of the DRP. These measurements were compared to observations obtained with the NIRC2 instrument on Keck II, an instrument with an excellent astrometric calibration, allowing us to derive an updated plate scale and north offset angle for GPI. This revised astrometric calibration should be used to calibrate all measurements obtained with GPI for the purposes of precision astrometry.
\end{abstract}


{\noindent \footnotesize\textbf{*}\linkable{rderosa@stanford.edu} }

\begin{spacing}{1}   

\section{Introduction}
\label{sec:intro}
The Gemini Planet Imager \cite{Macintosh:2008dn,Macintosh:2014js} (GPI) is an instrument, currently at the Gemini South telescope, Chile, that was designed to achieve high contrast at small angular separations to resolve planetary-mass companions around nearby, young stars. Many high-contrast imaging observations also require highly precise and accurate astrometry. One of the objectives of the large Gemini Planet Imager Exoplanet Survey\cite{Nielsen:2019cb} (GPIES) was to characterize via relative astrometry the orbits of the brown dwarfs and exoplanets imaged as a part of the campaign\cite{Konopacky:2014hf}. These measurements have been used to investigate the dynamical stability of the multi-planet HR 8799 system\cite{Wang:2018fd}, the interactions between substellar companions and circumstellar debris disks\cite{MillarBlanchaer:2015ha, Rameau:2016dx}, and to directly measure the mass of $\beta$~Pictoris~b \cite{Nielsen:td}. Improved astrometric accuracy and precision can reveal systematic discrepancies between instruments which need to be considered when performing orbital fits using astrometric records from multiple instruments. Accurate, precise astrometry can also help with common proper motion confirmation or rejection of detected candidate companions. 

Previous work has demonstrated that the location of a faint substellar companion relative to the host star can be measured within a reduced and post-processed GPI image to a precision of approximately seven hundredths of a pixel\cite{Wang:2016gl}. Since GPI's science camera is an integral field spectrograph/polarimeter, ``pixel'' in this context means the spatial pixel sampling set by the IFS lenslet array, rather than of the subsequent Hawaii-2RG detector. Converting these precise measurements of the relative position of the companion from pixels into an on-sky separation and position angle require a precise and accurate astrometric calibration of the instrument. The plate scale of the instrument is required to convert from pixels in the reconstructed data cubes into arcseconds, and the angle of north on an image that has been derotated to put north up based on the astrometric information within the header. The previous astrometric calibration (a plate scale of $14.166\pm0.007$\,mas\,px$^{-1}$ and a north offset angle of $-0.10\pm0.13$\,deg) was based on observations of calibration binaries and multiple systems obtained during the first two years of operations of the instrument \cite{Konopacky:2014hf, DeRosa:2015jl}.

In the course of several investigations using GPI that relied upon astrometric measurements, over time it became apparent that there were potentially remaining systematic biases after that calibration, particularly in regards to the north angle correction. This motivated a careful, thorough calibration effort into GPI astrometry, an effort that eventually grew to include cross checks of the GPI data processing pipeline, the performance of several Gemini observatory systems, and a complete reanalysis of all astrometric calibration targets observed with GPI. 

This paper presents the findings of those efforts, and the resulting improved knowledge of GPI's astrometric calibration. After introducing some background information regarding GPI and the Gemini architecture (Sec.~\ref{sec:architecture}), we describe two issues that we identified and fixed in the data reduction pipeline (Sec.~\ref{sec:updates}), a retroactive calibration of clock biases affecting some GPI observations (Sec.~\ref{sec:clocks}), and a model to calibrate for small apparent position angle changes in some observations at small zenith distances (Sec.~\ref{sec:rotator}). With those issues corrected, we revisit the astrometric calibration of GPI based on observations of several calibration binaries and multiple systems (Sec.~\ref{sec:new_calib} and \ref{sec:new_tn}). Compared to the prior calibration values, we find no significant difference in the plate scale. However we find a different value for the true north correction by +0.36 degrees, along with tentative low-significance evidence for small gradual drifts in that correction over time. Finally, we discuss the effect of the revised astrometric calibration on the astrometric measurements of several substellar companions (Sec.~\ref{sec:companion_astro}).

\section{GPI and Gemini Systems Architecture Context}
\label{sec:architecture}
\subsection{GPI Optical Assemblies}
\label{sec:architecture_optical}
\begin{figure}
\begin{center}
\begin{tabular}{c}
\includegraphics[width=16cm]{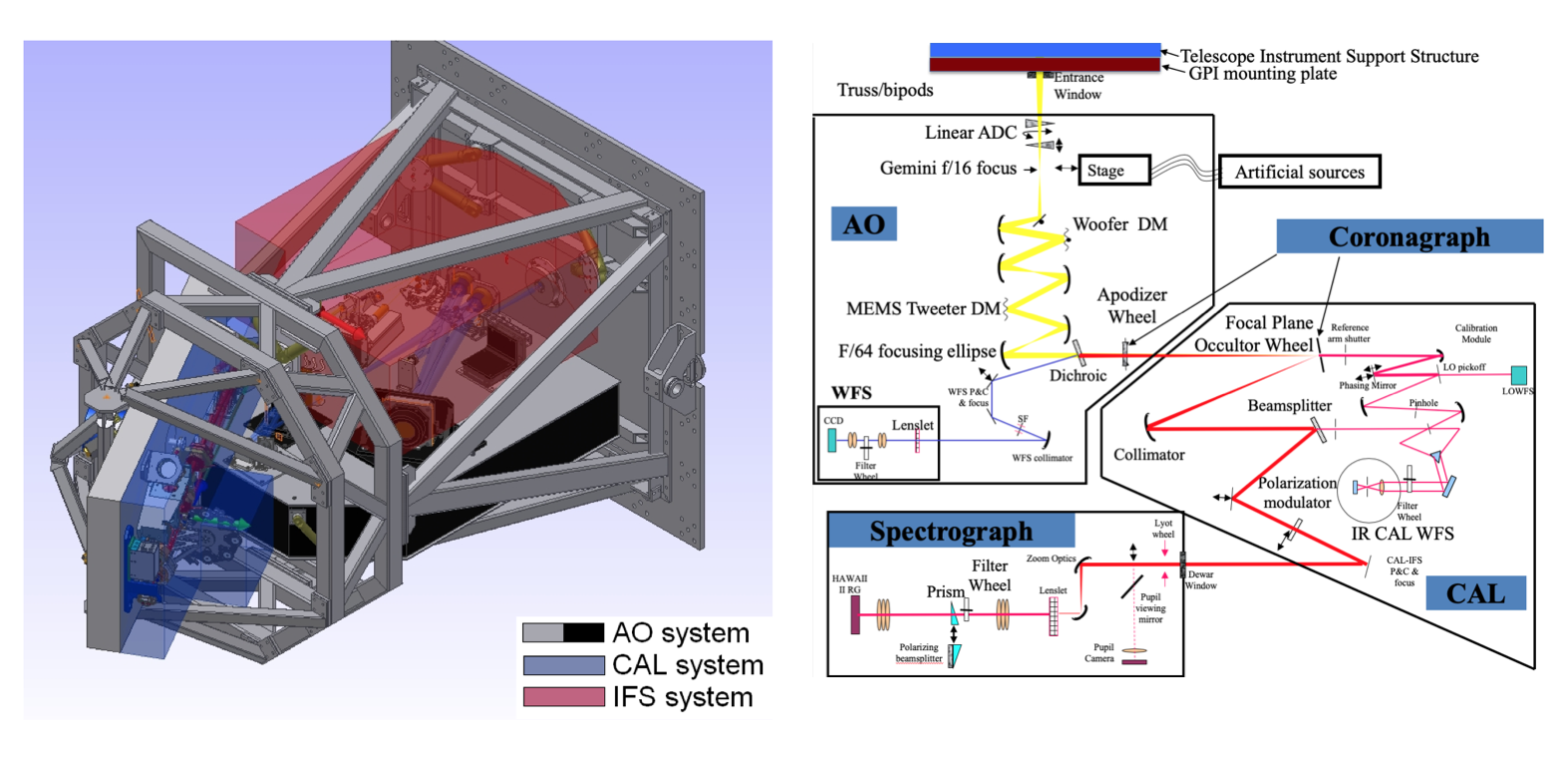}
\end{tabular}
\end{center}
\caption 
{ \label{fig:gpi_diagram}
Left: CAD rendering of the GPI assembly showing the AO, CAL, and IFS optical benches and the supporting truss structure and mounting plate. For scale, the mounting plate is 1.2 m on a side. (Note that this shows an earlier version of the truss, the as-built structure is slightly different.) Right: Schematic showing the light path through the three optical assemblies.} 
\end{figure}

The Gemini Planet Imager\cite{Macintosh:2008dn,Macintosh:2014js} combines three major optical assemblies (Fig.~\ref{fig:gpi_diagram}). The adaptive optics (AO) system is mounted on a single thick custom optical bench. The Cassegrain focus of the telescope is located within the AO assembly. On that bench, the beam encounters a linear thin-plate atmospheric dispersion corrector, steerable pupil-alignment fold mirror, an off-axis parabolic (OAP) relay to the first deformable mirror, and an OAP relay to the second deformable mirror. After that, the beam is refocused to f/64. The last optic on the AO bench is a wheel containing microdot-patterned coronagraphic apodizer masks\cite{Soummer:2009el,Soummer:2011eq}.  These apodizer masks also include a square grid pattern that induces a regular pattern of diffracted copies of the stellar point spread function\cite{Marois:2006kh,Sivaramakrishnan:2006cf} (PSF).

The second optical assembly is an infrared wavefront sensor known as the CAL system\cite{Wallace:2010hx}. It contains the focal plane mask component of the coronagraph (a flat mirror with a central hole), and collimating and steering optics. 

The third assembly is the integral field spectrograph\cite{Chilcote:2012hd,Larkin:2014ek} (IFS). The input collimated beam is refocused onto a grid of lenslets that serve as the image focal plane of the system. After this, the spectrograph optics relay and disperse the lenslet images, but since the beam has been segmented, these can no longer introduce astrometric effects. The lenslet array samples the focal plane and produces a grid of “spots” or micropupils which are each an image of the telescope pupil. The only aberrations affecting the image quality of the field are from elements in front of the lenslet array\cite{Larkin:2014ek}.

Each of these three assemblies is independently mounted by three bipods. The bipods are supported by a steel truss structure that attaches to a square front mounting plate. The mounting plate attaches to the Gemini Instrument Support Structure (ISS) with large fixed kinematic pins. The ISS is a rotating cube located just above the Cassegrain focus of the telescope.

In typical Gemini operations, the ISS rotator operates to keep the sky position angle fixed on the science focal plane. High-contrast imaging typically instead tries to fix the telescope pupil on the science instrument to allow angular differential imaging\cite{Marois:2006df} (ADI). In GPI's case, this is always done at a single orientation (corresponding to GPI's vertical axis parallel to the telescope vertical axis.) In the simplest case, this would involve stopping all rotator motion. However, as discussed in Section~\ref{sec:rotator}, in some but not all observations the observatory software instead tries to maintain the absolute (sky) vertical angle stationary on the science focal plane, which must be accounted for in astrometric observations. 

\subsection{Software Interface and IFS Operation}
The software architecture for GPI and the Gemini South telescope is complex, as typical for a major observatory. Simple operations often require interactions between several different computers. For example, taking an image with the IFS is a process that involves four separate computer systems; the main Gemini environment which runs the observatory's control software, GPI's top level computer (TLC) that is interfaced with each component of the instrument, the IFS ``host'' computer that acts as an interface between the UNIX-based TLC and the Windows-based detector software, and the IFS ``brick'' that interfaces directly with the Hawaii-2RG detector\cite{Larkin:2014ek}. Three of these four computer systems are responsible for populating the Flexible Image Transport System\cite{2010A&A...524A..42P} (FITS) image header keywords appended to each image. The Gemini environment handles telescope-specific quantities such as the telescope mount position, the TLC handles keywords associated with other parts of the instrument such as the AO system, and the IFS brick records detector-specific quantities. Each of these computer systems also maintains its own clock, although only the clock of the Gemini and environment and the IFS brick are relevant for the purposes of this study. These clocks are used when appending various timestamps to FITS headers during the process of obtaining an image. In theory, these clocks should all be synchronized periodically with Gemini's Network Time Protocol (NTP) server.

The IFS camera is controlled by the IFS brick, a computer used to interface with the Teledyne JADE2 electronics and Hawaii-2RG detector. This computer is responsible for commanding the camera, calculating count rates for each pixel based on raw up-the-ramp (UTR) reads\cite{Fowler:1991gq}, sending completed images back to the observatory computers, and providing ancillary metadata including the start and end time of the exposure ({\tt UTSTART} and {\tt UTEND}) that are stored in the FITS header. The detector is operated almost exclusively in UTR mode; correlated double sampling (CDS) mode\cite{McLean:2008ua} images have been taken in the laboratory, but this mode is not available for a standard observing sequence. The IFS runs at a fixed pixel clocking rate of 1.45479\,s for a full read or reset of the detector. The IFS software allows for multiple exposures to be coadded together prior to writing a FITS file. This mode has lower operational overheads and greater operational efficiency compared to individual exposures, and therefore is frequently used for short exposures (from 1.5 to 10\,s per coadd), but not generally used for long exposures (60\,s per coadd) due to field rotation.

\section{Improvements in the GPI Data Reduction Pipeline}
\label{sec:updates}
The GPI Data Reduction Pipeline\cite{Perrin:2014jh,Perrin:2016gm} (DRP) is an open-source pipeline that performs basic reduction steps on data obtained with GPI's IFS, to remove a variety of instrumental systematics and produce science-ready spectrophotometrically- and astrometrically-calibrated data cubes. The DRP corrects for detector dark current, identifies and corrects bad pixels and cosmic ray events, extracts the microspectra in the 2-D image to construct a 3-D ($x,y,\lambda$) data cube (or $x,y,\mathrm{Stokes}$ in polarimetry mode), and corrects for the small geometric distortion measured in the laboratory during the integration of the instrument\cite{Konopacky:2014hf}. 

\begin{figure}
\begin{center}
\begin{tabular}{c}
\includegraphics[width=10cm]{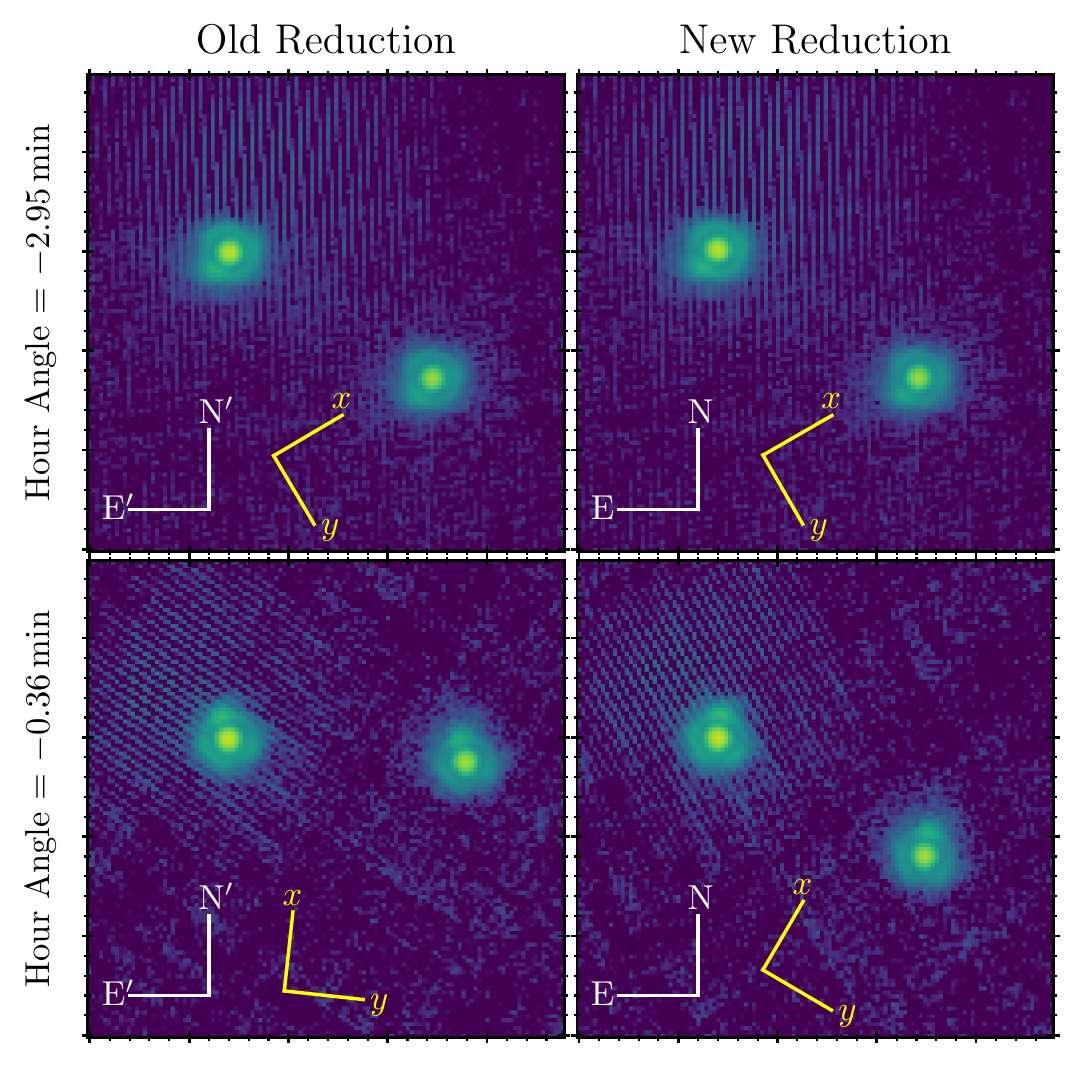}
\end{tabular}
\end{center}
\caption 
{ \label{fig:hd6307-example}
Two GPI images of the binary star HD 6307 demonstrating the error in the calculation of {\tt AVPARANG} in the previous version of the pipeline. The old reduction (left column) and new reduction (right column) for two images approximately three minutes (top row) and less than one minute (bottom row) before the target transited the meridian. Each image has been rotated such that North is up based on the value of {\tt AVPARANG} in the header of the reduced image (white compass). We use the prime symbol to denote the fact that the old reduction does not correctly rotate North up. The original detector coordinate axes are also shown (yellow compass). Note the flip of the $x$-axis due the odd number of reflections within the instrument. A significant change in the sky position angle of the companion is seen between the two images in the left column, due to a combination of the errors described in this section. The position angle of the companion is stable after the revisions to the pipeline.} 
\end{figure}

Critically, the DRP calculates the average parallactic angle between the start and end of an exposure, an angle that is used to rotate the reduced data cubes so that the vector towards celestial north is almost aligned with the columns of the image. We have identified, and corrected in the latest data pipeline version, two issues with the calculation of average parallactic angle which affect a subset of GPI measurements. These issues are most pronounced for observations taken at a very small zenith distance, where the parallactic angle is changing very rapidly. An example dataset showing the combined effect of these issues, and those described in Sections \ref{sec:clocks} and \ref{sec:rotator}, on observations of a calibration binary is shown in Figure~\ref{fig:hd6307-example}.

\subsection{Calculation of Average Parallactic Angle from Precise Exposure Start and End Times}

Calculating the time-averaged parallactic angle during the course of an exposure requires accurate and precise knowledge of the exact start and end times of that exposure. We found that the GPI DRP was not originally using a sufficiently precise value for the start time in the case of an exposure with more than one coadd. Doing this correctly requires an understanding of the low-level details of the up-the-ramp readout of the Hawaii-2RG detector and the surrounding GPI and Gemini software. 

\begin{figure}
\begin{center}
\begin{tabular}{c}
\includegraphics[width=16cm]{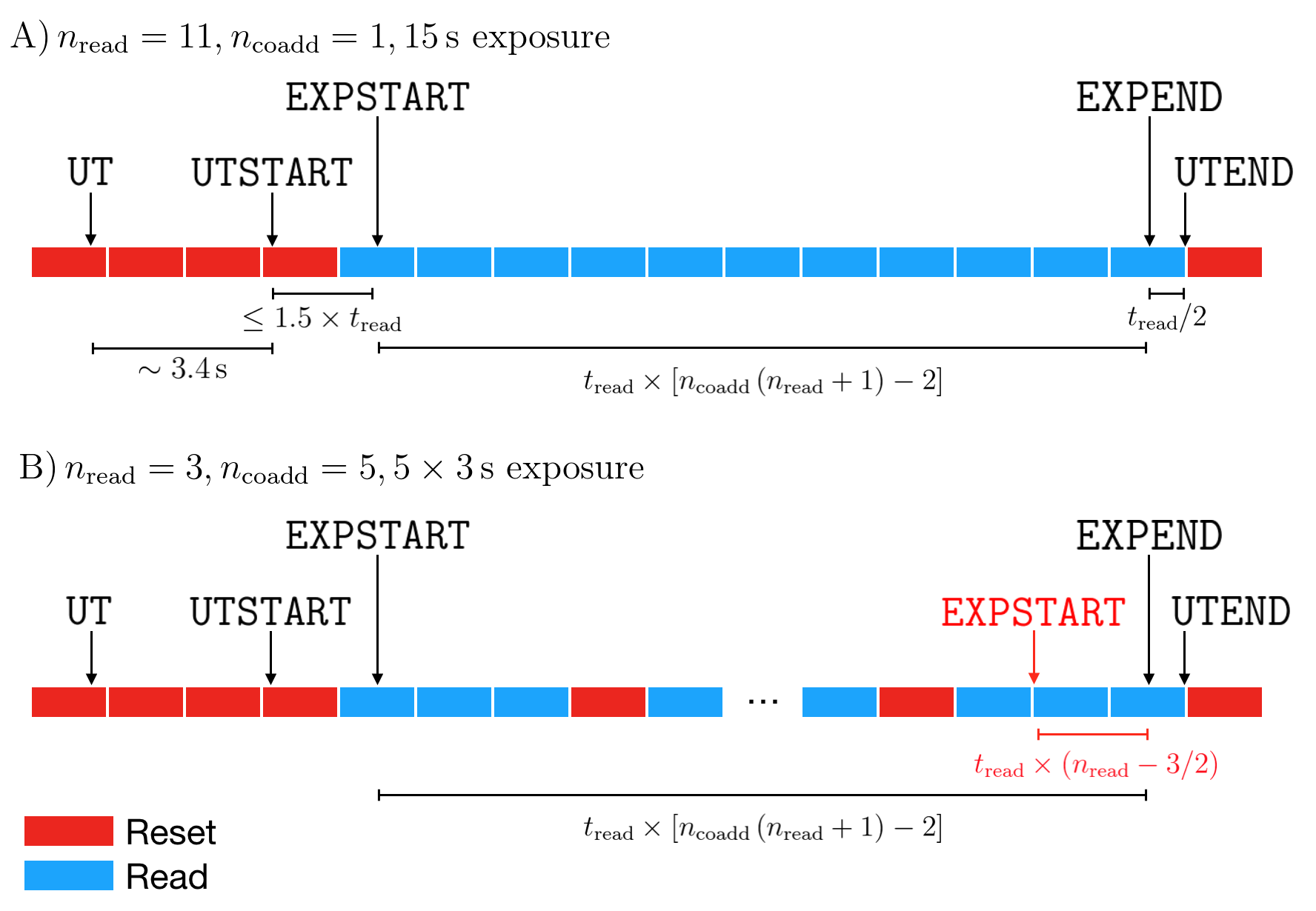}
\end{tabular}
\end{center}
\caption 
{ \label{fig:ifs_reads}
Reads (blue) and resets (red) of the Hawaii-2RG for two example exposures: a single coadd exposure with 11 reads (top), and a five coadd exposure with three reads per coadd (bottom). The Hawaii-2RG is in continuous reset mode prior to the start of an exposure. The {\tt UTSTART} keyword is generated when the exposure is commanded by the IFS software, which can be up to one and a half times the read out time prior to the start of the exposure. {\tt UTEND} is generated at the end of the final read. The {\tt EXPSTART} and {\tt EXPEND} values are calculated by the pipeline. The erroneous formula for computing {\tt EXSPTART} for exposures with coadds is shown in red in the bottom panel.} 
\end{figure}

The header of a raw GPI FITS file contains four timestamps saved at various times during the acquisition of an image with the IFS: {\tt UT}, {\tt MJD-OBS}, {\tt UTSTART}, and {\tt UTEND}. The keywords {\tt UT} and {\tt MJD-OBS} contains the time at the moment the header keyword values were queried by the Gemini Master Process (GMP) prior to the start of the exposure.  {\tt UT} is reported in the Coordinated Universal Time (UTC) scale, whereas {\tt MJD-OBS} is reported in the Terrestrial Time scale, a scale linked to International Atomic Time that is running approximately 65 seconds ahead of UTC. Because these keywords are written during exposure setup by a different computer system, neither is a highly precise metric for the exact exposure time start. The other keywords ({\tt UTSTART}, {\tt UTEND}) are generated by the IFS brick upon receipt of the command to execute an exposure, and after the final read of the last coadd has completed. These two timestamps are reported in the UTC scale. Because they are written by the same computer that directly controls the readout, these are more accurate values for exposure timing. {\tt UTSTART} is written when the IFS software receives the command to start an exposure, but since the Hawaii-2RG will be in continuous reset mode between exposures, it must wait some fraction of a read time to complete the current reset before the requested exposure can begin. Thus the true exposure start time will be some unknown fraction of a read time after {\tt UTSTART}. The final keyword {\tt UTEND} is written with negligible delay immediately at the moment the last read of the last pixel is concluded. A schematic diagram of the read and resets of the Hawaii-2RG is shown for two example exposures in Figure~\ref{fig:ifs_reads}.

The pipeline was therefore written under the assumption that the {\tt UTEND} keyword provides the most accurate way to determine the true start and end time of each exposure, which in turn is used to calculate the average parallactic angle during the exposure. The effective end time of the exposure can be calculated as occurring half a read time prior to {\tt UTEND}, i.e. the time at which half of the detector pixels have been read. The effective start time of the exposure, i.e. when half of the detector pixels have been read for the first time, can be calculated working backwards from {\tt UTEND} towards {\tt UTSTART}.
We do so based on the read time ($t_{\rm read}$), number of reads per coadd ($n_{\rm read}$, where $n_{\rm read} -1$ multiplied by $t_{\rm read}$ yields the integration time per coadd), and number of coadds ($n_{\rm coadd}$). The pipeline writes two additional keywords to the science extension of the reduced FITS file that store the calculated effective start ({\tt EXPSTART}) and end ({\tt EXPEND}) times of the exposure calculated using {\tt UTEND}, $t_{\rm read}$, $n_{\rm read}$, and $n_{\rm coadd}$. {\tt EXPSTART} and {\tt EXPEND} are then used to calculate the average parallactic angle over the course of the exposure, which is written as keyword {\tt AVPARANG}.

\begin{figure}
\begin{center}
\begin{tabular}{c}
\includegraphics[width=10cm]{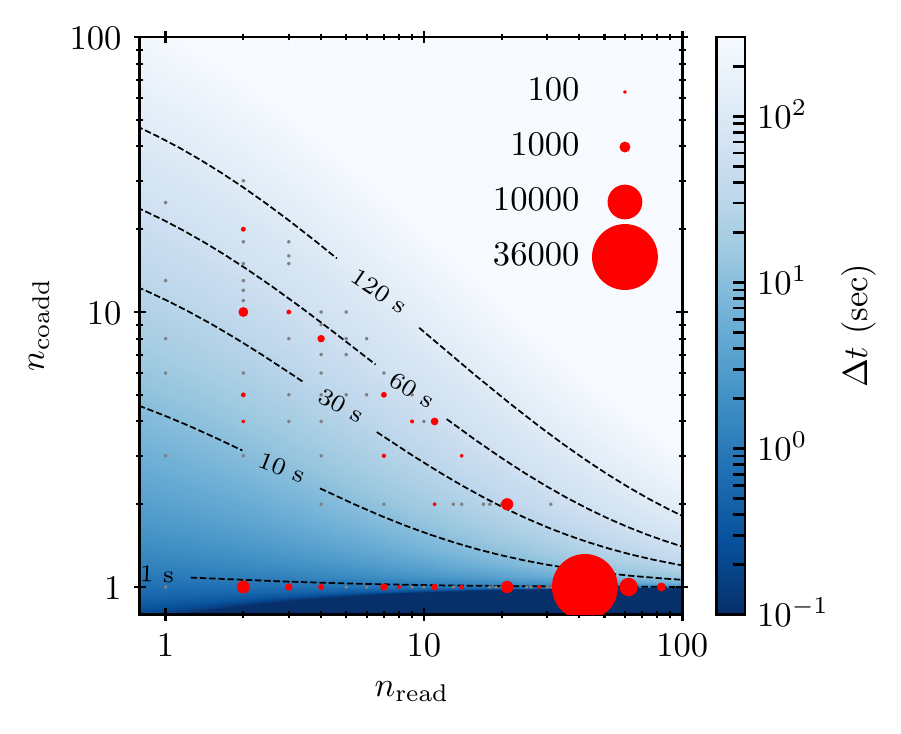}
\end{tabular}
\end{center}
\caption 
{ \label{fig:deltat}
Error in the calculated duration of an exposure as a function of the number of reads (approximately equivalent to the integration time per coadd divided by 1.45\,s) and the number of coadds. Dashed lines denote contours of $\Delta T = 1$, 10, 30, 60, and 120\,s. All unique combinations of $n_{\rm read}$ and $n_{\rm coadd}$ for all on-sky GPI images within the GPIES database are plotted. Combinations with more than 100 images are shown as red circles (size scaled by the number), while combinations with less than 100 are shown as small gray circles. The vast majority of GPI exposures are taken with a single coadd, but for some frames with multiple coadds $\Delta T$ exceeded 120\,s.} 
\end{figure}

Inadvertently, versions 1.4 and prior of the GPI pipeline contained an error in this calculation by not correctly accounting for the number of coadds. The total exposure time including overheads was calculated as $t_{\rm exp} = t_{\rm read}\times \left(n_{\rm read}-3/2\right)$, where $n_{\rm read}$ is the number of reads {\it per coadd}. Instead, the exposure time is more correctly calculated as
\begin{equation}
t_{\rm exp} = t_{\rm read}\times (n_{\rm coadd} \times (n_{\rm read} +1) - 2),
\end{equation}
where the additional terms account for the extra resets that occur between each coadd. The effect of this error was negligible for single-coadd exposures, the most common type of exposures taken with GPI; 89\% of on-sky observations were taken with a single coadd. For images with multiple coadds the effect can be very significant, with the error on the estimated time elapsed during the complete observation of
\begin{equation}
    \Delta t = t_{\rm read}\times \left(n_{\rm coadd} \times n_{\rm read} + n_{\rm coadd} - n_{\rm read} - 1/2\right).
\end{equation}
To demonstrate how large this error can get for exposures with multiple coadds, an exposure with an integration time of 1.45 seconds with ten coadds has a $\Delta t$ of 40 seconds, an error equivalent to 98\% of the actual time spent exposing (see Fig. \ref{fig:deltat}). A large $\Delta t$ can cause a significant and systematic error in the parallactic angle used to rotate the reduced data cubes north up as {\tt EXPSTART} and {\tt EXPSTOP} header keywords are converted into the hour angle at the start and end of the exposure from which the parallactic angle is calculated. This is most pronounced for targets observed at a small zenith distance where the parallactic angle is changing most rapidly. This error not only affects astrometry of substellar companions, but also the measurement of binaries observed with other instruments that were used to calibrate GPI's true north offset angle.

After this inaccuracy was discovered, the GPI pipeline was updated to perform the correct calculation, as of version 1.5. 

\subsection{Average Parallactic Angle During Transits}
\label{sec:romberg}
A second issue affecting a small number of observations is related to time-averaging during exposures that span transit. 

The pipeline computes the average parallactic angle between the start and end of an exposure via Romberg's method. For northern targets that transit during an exposure, the function contains a discontinuity at an hour angle ($H$) of $H=0$\,rad where the parallactic angle jumps from $-\pi$ to $+\pi$. This discontinuity can be easily avoided by performing the integration between $H=H_0$ and $H=0$\,rad, and between $H=0$\,rad and $H=H_1$, where $H_0$ and $H_1$ are the hour angle at the start and end of the exposure. The prior version 1.4 of the pipeline and earlier contained an error in how this calculation was performed. As an example, the average parallactic angle $p_{\rm avg}$ for an exposure with $|H_0| < H_1$ was calculated as
\begin{equation}
    p_{\rm avg} =\frac{1}{H_1-H_0}\left[\int_{H_0}^0 |p\left(H, \phi,\delta\right)| dH + \int_0^{H_1} p\left(H, \phi,\delta\right) dH\right]
\end{equation}
rather than 
\begin{equation}
    p_{\rm avg} = \frac{1}{H_1-H_0}\left[\int_{H_0}^0 \left[p\left(H, \phi,\delta\right) + 2\pi \right]dH + \int_0^{H_1} p\left(H, \phi,\delta\right) dH  \right].
\end{equation}
This error only affects sequences where the target star transited the meridian between the start and end of an exposure. The magnitude of this error depends on exactly when transit occurred relative to the start and the end, and the declination of the target. The net effect of the error on companion astrometry is small as it will only affect one of approximately forty images taken in a typical GPI observing sequence.

This issue has also been corrected as of the latest version of the GPI pipeline. 

\section{Inaccuracies in Some FITS Header Time Information}
\label{sec:clocks}
The pipeline necessarily relies on the accuracy of the FITS header keywords in the data it is processing; however it has proven to be the case that the FITS header keyword time information is not always as reliable as we would like. A review of FITS header timing information allowed us to uncover several periods in which misconfiguration or malfunction of time server software resulted in systematic errors in header keyword information. We were able to reconstruct the past history of such timing drifts sufficiently well as to be able to retroactively calibrate it out when reprocessing older data. 

\begin{figure}
\begin{center}
\begin{tabular}{c}
\includegraphics[width=10cm]{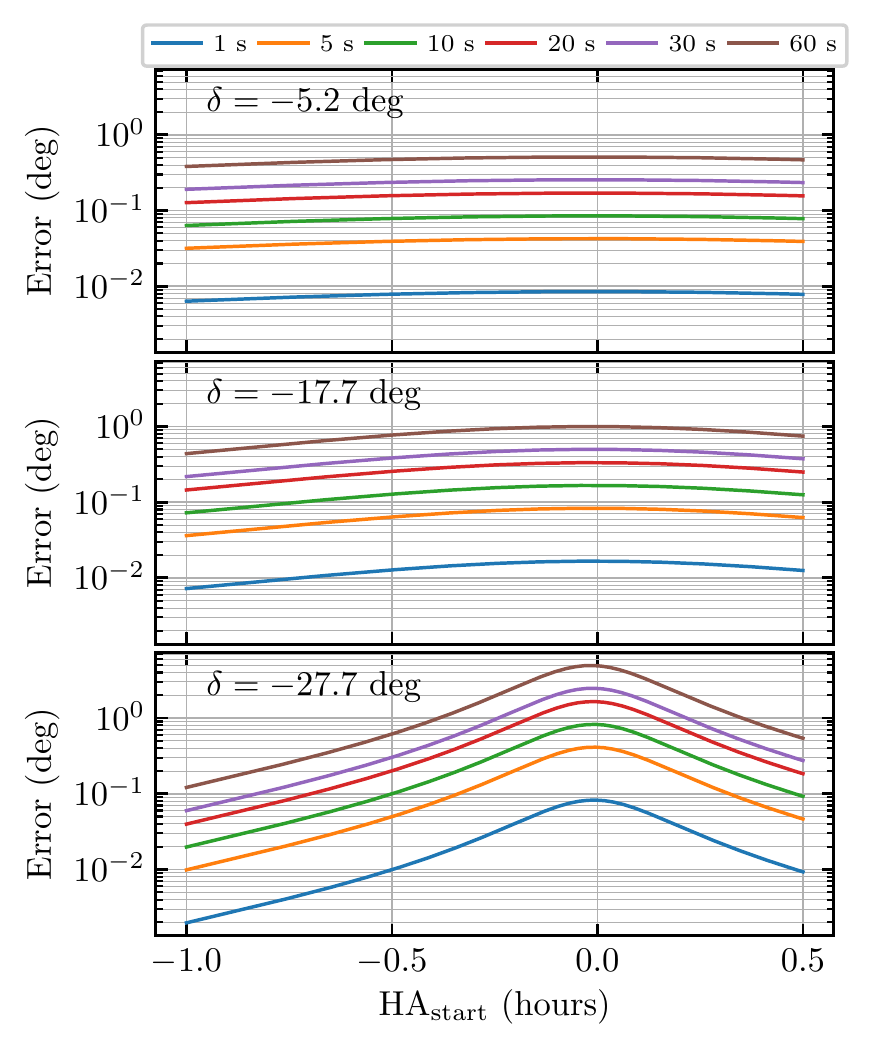}
\end{tabular}
\end{center}
\caption 
{ \label{fig:parang_error}
Error in the value of {\tt AVPARANG} for a 60\,s exposure induced by an IFS clock drift of between 1\,s and 60\,s as a function of the hour angle at the start of the exposure for three target declinations. The error is most significant for targets that transit the meridian with a small zenith distance (bottom panel).} 
\end{figure}

As a reminder the {\tt UTSTART} keyword is written by the IFS brick computer. The clock on the IFS brick is, at least in theory, configured to automatically synchronize once per week with Gemini's NTP server. This server provides a master time reference signal to maintain the accurate timings necessary for telescope pointing and control. In order to cause a noticeable error in the average parallactic angle, the IFS brick time stamps would have to be between a few and a few tens of seconds out of sync, depending on the declination of the star (Fig.~\ref{fig:parang_error}). The regular synchronization of the clock on the IFS brick was intended to be sufficient to prevent it from drifting at such an amplitude relative to the time maintained by Gemini's NTP server. 

However, it was eventually discovered that this time synchronization has not always operated as intended, resulting in significant clock offsets for some periods. The history of the offset between the IFS brick clock and UTC cannot be recovered directly from the various logs and headers generated by the IFS. Instead, we can use the difference between the {\tt UT} and {\tt UTSTART} header keywords as a proxy. The first timestamp is generated when the command to execute an observation is issued by Gemini's Sequence Executor (SeqExec) and is assumed to be accurate; a significant offset in the observatory's clock would quickly become apparent when attempting to guide the telescope. The second timestamp is generated when the IFS brick receives the command to start an exposure from the GPI Top Level Computer (TLC). The difference between these two timestamps, {\tt UTSTART-UT}, should be small and relatively stable, as there has not been any significant changes to these software components since the instrument was commissioned in 2014, and we show below that this time difference does prove to be stable for the majority of GPI data.

We therefore data mined all available GPI data to determine the time evolution of the offset between {\tt UT} and {\tt UTSTART} during the entire time GPI has been operational. We queried the GPIES SQL database\cite{Wang:2017ju,2018SPIE10703E..0HS}, which contains the header information for all images obtained in the GPIES Campaign programs, selected guest observer (GO) programs whose PIs have contributed their data into this database, and all public calibration programs. We augmented this with all GO programs that were publicly accessible in the Gemini Observatory Science Archive when this analysis was performed. We excluded engineering frames---images that are obtained via GPI's IDL interface---as the {\tt UT} keyword is populated via a different process for these types of frames. A total of 99,695 measurements of the {\tt UT} to {\tt UTSTART} offset spanning the previous six years were obtained, including 93,575 from the GPIES database and 6,120 from other GO programs not included within the database. 

\begin{figure}
\begin{center}
\begin{tabular}{c}
\includegraphics[width=16cm]{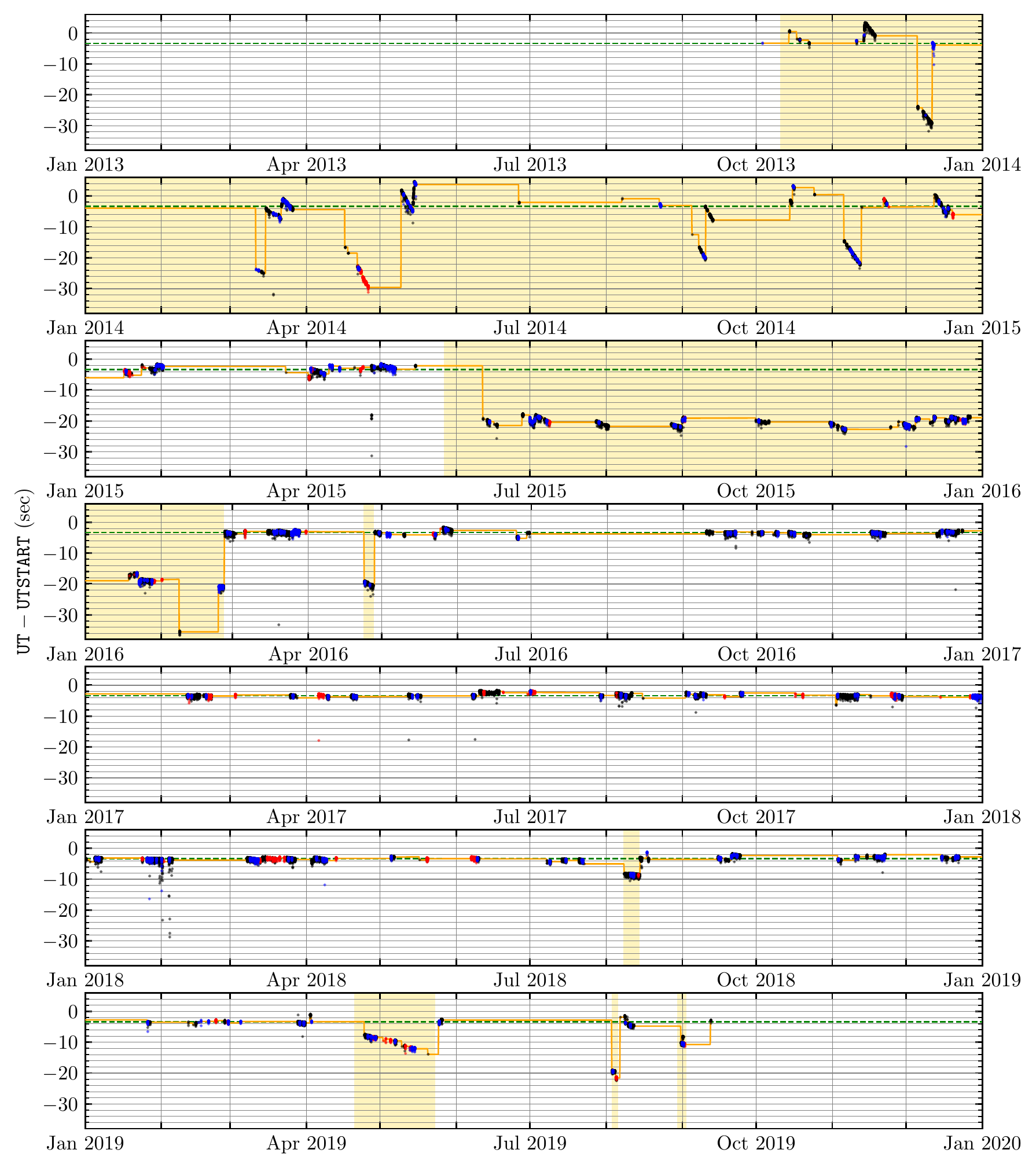}
\end{tabular}
\end{center}
\caption 
{ \label{fig:ut_offset}
Difference between {\tt UT} and {\tt UTSTART} in the header of coronagraphic (black), polarimetric (blue) images taken during the GPIES campaign, and for guest observer images (red). The green dashed line denotes the nominal offset between these keywords of 3.38\,s (see Fig.~\ref{fig:ref_offset}). The median offset calculated as the median of all frames within a 12-hour window (yellow solid line) was used to identify dates where the clock drift was significant (yellow shaded region).} 
\end{figure}

The evolution of this offset between the installation of the instrument at Gemini South and now is shown in Figure~\ref{fig:ut_offset}. We identified several periods of time, two quite extended, where the IFS clock was not correctly synchronized with the Gemini NTP. From the initial commissioning of the instrument until the end of 2014 the offset varied significantly, from about eight seconds slow to up to thirty seconds fast. The causes of these variations are not fully known, but we point out that during this first year, GPI was still in commissioning and shared-risk science verification, and software was still significantly in flux. In several instances, negative shifts in the offset are correlated with dates on which the IFS brick was used after having been restarted but prior to the periodic time synchronization having occurred.  The gradual negative drifts in offset observed at several points imply that the IFS clock was running too fast, gaining time at a rate of approximately one second per day over this period.  Later, other small excursions in April 2016, August 2018, and August 2019 were also apparently caused by the IFS brick being used after an extended time powered off but prior to the scheduled weekly time synchronization. It would of course have been better had the time synchronization occur automatically immediately after each reboot, but that was not the case.

A second long period with a significant offset, between June 2015 and March 2016, was caused by the IFS brick being synchronized to the wrong time server; it was tracking the GPS time scale rather than UTC, and therefore ran 18 seconds ahead of UTC.  An extended drift in the offset from April into May 2019 was caused by a failure in the NTP daemon running on a computer intermediate to the IFS brick and Gemini's NTP server. The drift was noticeably slower than in the 2013-2014 period, with the IFS brick gaining time at a rate of only one quarter of a second per day.

\begin{figure}
\begin{center}
\begin{tabular}{c}
\includegraphics[width=10cm]{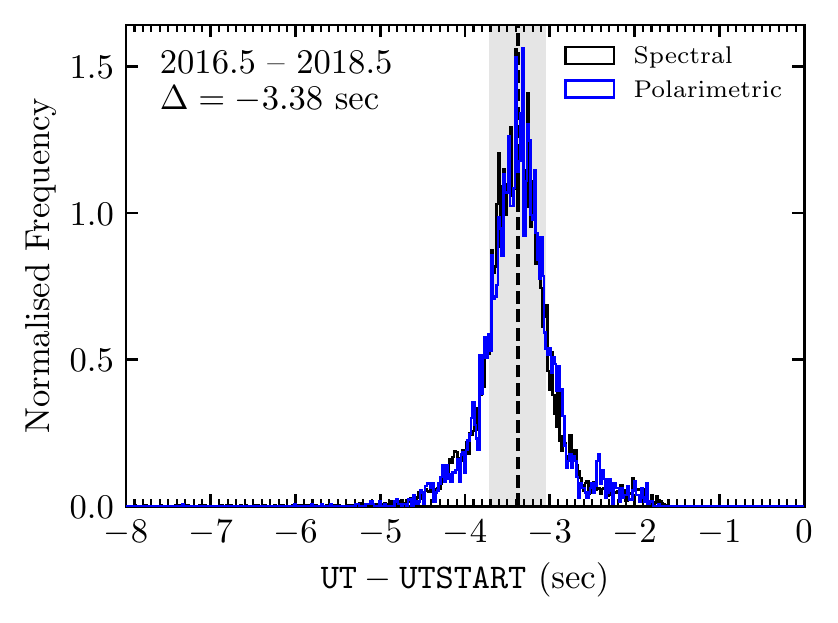}
\end{tabular}
\end{center}
\caption 
{ \label{fig:ref_offset}
Histogram of the offset between {\tt UT} and {\tt UTSTART} for spectroscopic (black) and polarimetric (blue) observations taken between 2016.5 and 2018.5, when there were no clock biases. The width of the distribution is narrow relative to the 20 to 30 s offsets shown in the prior figure, supporting the notion that we can use drifts in {\tt UT-UTSTART} to track the clock biases affecting {\tt UTSTART} and {\tt UTEND}.} 
\end{figure} 

Improved systems administration can prevent such drifts in the future, but in order to properly calibrate the available data we must model out the drifts that occurred in the past. 
The offset between {\tt UT} and {\tt UTSTART} remained relatively stable from mid-2016 through mid-2018, and was independent of the observing mode. We measured the median offset value between 2016.5 and 2018.5 as $-3.38$\,sec and defined this as the nominal {\tt UT} to {\tt UTSTART} offset (Fig. \ref{fig:ref_offset}). We used a rolling median with a width of 12 hours to calculate the value of the offset at a resolution of one hour between late 2013 and 2019. A lookup table was created that the pipeline queries when reducing an IFS image, so that it can apply a correction to {\tt UTSTART} and {\tt UTEND} if the observation was taken during a period identified as having a significant offset (Fig. \ref{fig:ut_offset}).

\section{Modeling Apparent Image Rotation at Gemini's Cassegrain Port}
\label{sec:rotator}
\begin{figure}
\begin{center}
\begin{tabular}{c}
\includegraphics[width=10cm]{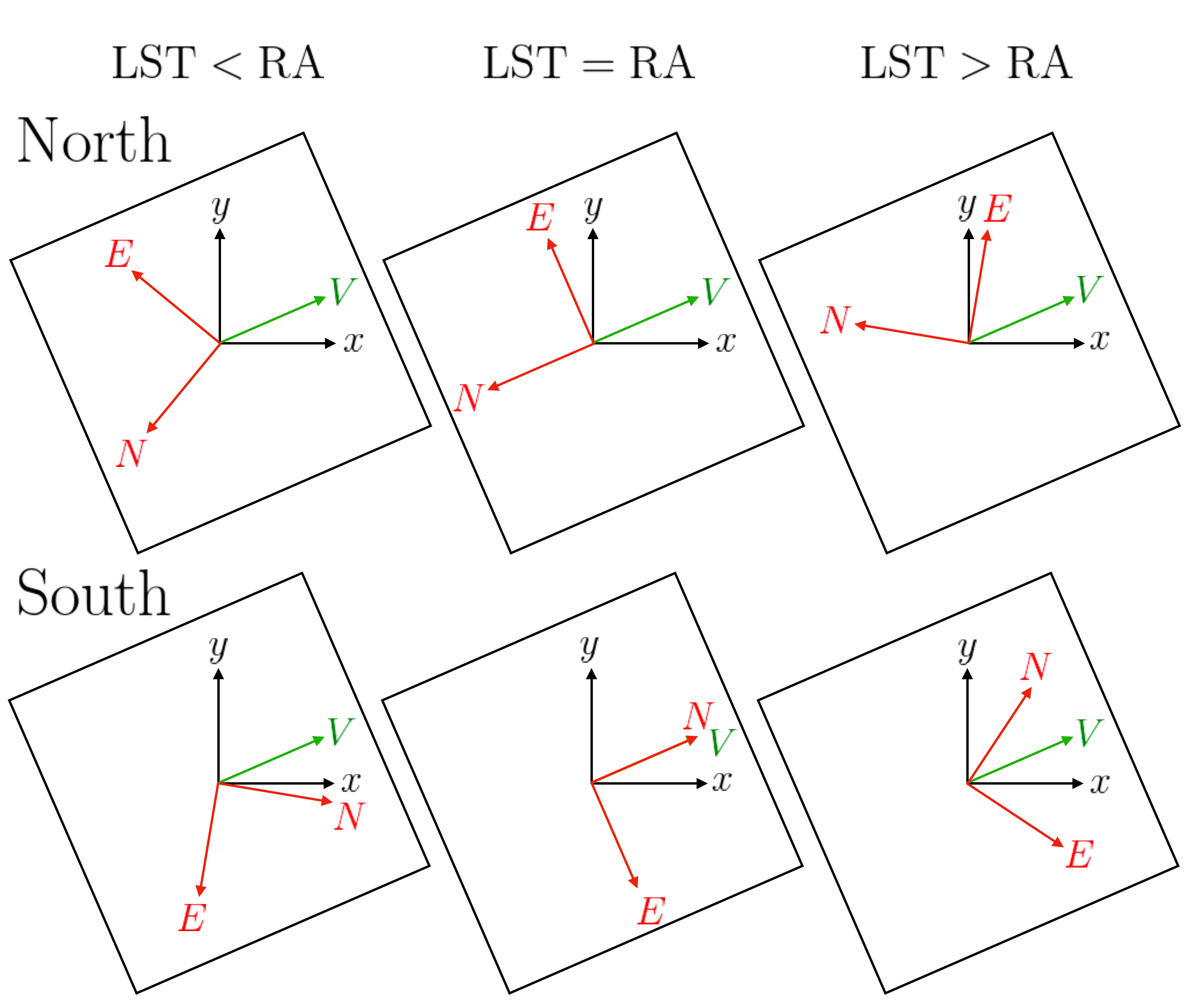}
\end{tabular}
\end{center}
\caption 
{\label{fig:gpi_coordinates}
Image (black) and sky (red) coordinate systems for observations taken before (left column) during (middle column) and after (right column) meridian transit for northerly (top row) and southerly (bottom row) targets. The angle of the vertical vector (green) remains fixed relative to the image coordinate system for an ideal altitude-azimuth telescope, here at an angle of approximately 23.5 degrees from the $x$-axis within a reduced GPI data cube. Any offset between true vertical and the vertical axis of the telescope will cause the vertical vector within a reduced image to move slightly as the target crosses the meridian, the magnitude of which would be imperceptible in this diagram for a small offset as is the case for Gemini South, but significant relative to the precision of astrometric measurements made with GPI.}
\end{figure}

\begin{figure}
\begin{center}
\begin{tabular}{c}
\includegraphics[width=10cm]{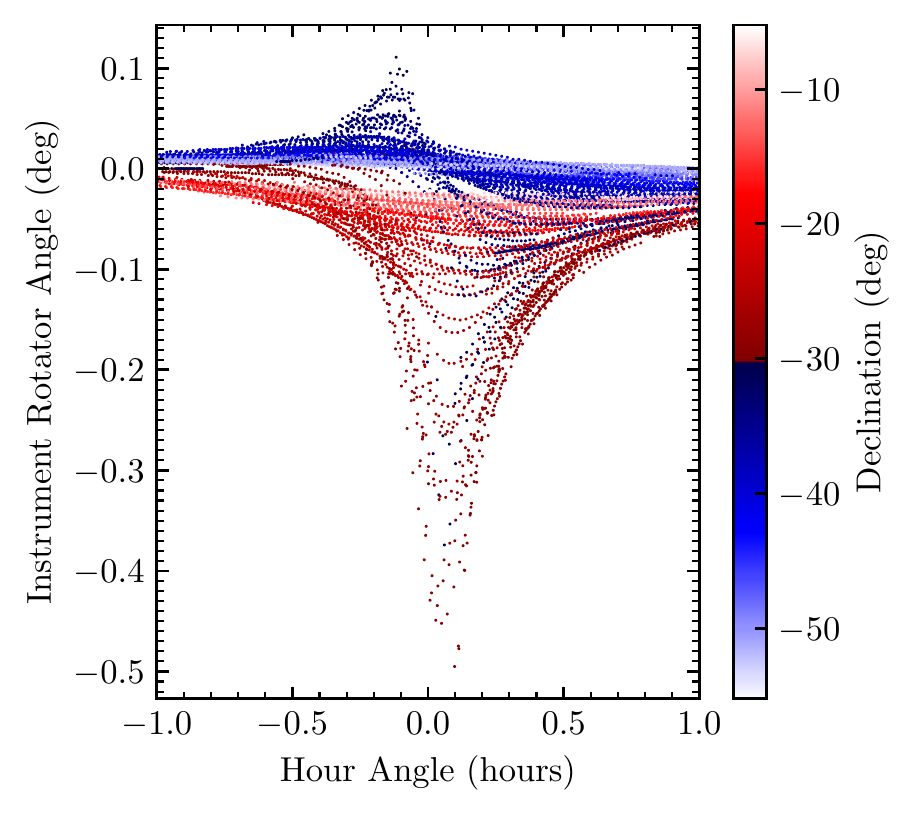}
\end{tabular}
\end{center}
\caption 
{ \label{fig:crpa}
Angle of the instrument rotator as a function of hour angle for GPI observations where the rotator drive was enabled. The color of the symbol denotes the declination of the target. The instrument rotator angle has a different behavior for northern and southern targets due to the non-perpendicularity of the Gemini South telescope.} 
\end{figure}

Recall from Section~\ref{sec:architecture_optical} that GPI always operates in ADI mode, with its pupil fixed or nearly fixed relative to the telescope pupil. GPI is attached to Gemini's ISS, which itself is mounted on the Cassegrain port of the telescope. A Cassegrain instrument rotator is used to maintain a fixed position angle between the columns on an instrument's detector and either celestial North or the zenith. For an ideal altitude-azimuth telescope with an elevation axis perfectly aligned with local vertical and with an azimuth platform perpendicular to vertical, an instrument mounted on the Cassegrain port would observe the North angle changing with the parallactic angle as the telescope tracked a star through the meridian. The angle between the columns on the instrument detector and the direction of vertical would remain fixed (Fig.~\ref{fig:gpi_coordinates}). Differences between true vertical and the vertical axis of the telescope cause this angle to vary slightly, an effect most pronounced for stars observed near the meridian with a small zenith distance ($\lesssim$5\,deg). When enabled, Gemini South's instrument rotator compensates for this motion, keeping the vertical angle fixed on the detector (Fig.~\ref{fig:crpa}). 

Due to difficulties maintaining the AO guide loops for targets with a very small zenith distance, it became common for some operators to keep the instrument rotator drive disabled while GPI was in operation, regardless of the target elevation. However, this practice was inconsistently applied. The drive was disabled and rotator kept at a nominal home position for 99 of the 317 nights on which GPI was used over the last six years. For data taken on these nights, a small correction needs to be applied to the parallactic angle in the header to compensate for this small motion of the vertical angle as a star is tracked through the meridian.

Such a correction relies on precise knowledge of the telescope mount alignment. Sufficiently precise information on the Gemini South telescope mount is not publicly available. We therefore derived \textit{post facto} knowledge of the Gemini South telescope mount based on the behavior of the Cassegrain rotator on nights when it was activated. 

We constructed a simple model to predict the correction to the parallactic angle caused by the non-perpendicular nature of the telescope \cite{Green:1985ve}. For a perfect telescope, the parallactic angle of a source $p$ is calculated as
\begin{equation}
\tan p = \frac{-\cos\phi\sin A}{\sin\phi \cos E - \cos\phi\sin E \cos A}
\end{equation}
where $A$ and $E$ are the topocentric horizontal coordinates of the target, i.e azimuth and elevation. If the telescope's azimuth platform is tilted at an angle of $\theta$ with an azimuth of $\Omega$, the difference between the true $p$ and apparent $p^{\prime}$ parallactic angles  is\footnote{\url{http://www.astro.caltech.edu/~mcs/CBI/pointing/}}
\begin{equation}
p^{\prime} - p = \Delta p = -\arctan\left(\frac{\cos \omega \sin \theta}{\cos E \cos \theta + \sin E \sin \theta \sin \omega}\right),
\end{equation}
where $\omega = \Omega-\pi/2-A$. A tilt in the elevation axis of $\theta_E$ within the plane connecting $A\pm\pi/2$ causes an additional modification of
\begin{equation}
\Delta p = -\arcsin(\sin \theta_E/\cos E)
\end{equation}
These tilts will lead to a slight difference in the elevation and azimuth ($E^{\prime}$, $A^{\prime}$) of the telescope mount versus the topocentric elevation and azimuth ($E$, $A$) of the target. The telescope elevation and azimuth modified by the azimuth tilt are calculated as 
\begin{equation}
\begin{split}
    \sin E^{\prime} &= \left(\sin E \cos \theta - \cos E \sin \theta \sin \omega
\right)\\
    A^{\prime} &= \Omega - \arctan\left(\frac{\cos\omega\cos E}{-\cos\theta \sin\omega\cos E - \sin\theta\sin E}\right),
\end{split}
\end{equation}
and due to an elevation tilt as
\begin{equation}
    \begin{split}
        \sin E^{\prime} &= \sin E/\cos \theta_E\\
        A^{\prime} &= A - \arcsin \left(\tan E \tan \theta_E\right).
    \end{split}
\end{equation}

\begin{figure}
\begin{center}
\begin{tabular}{c}
\includegraphics[width=16cm]{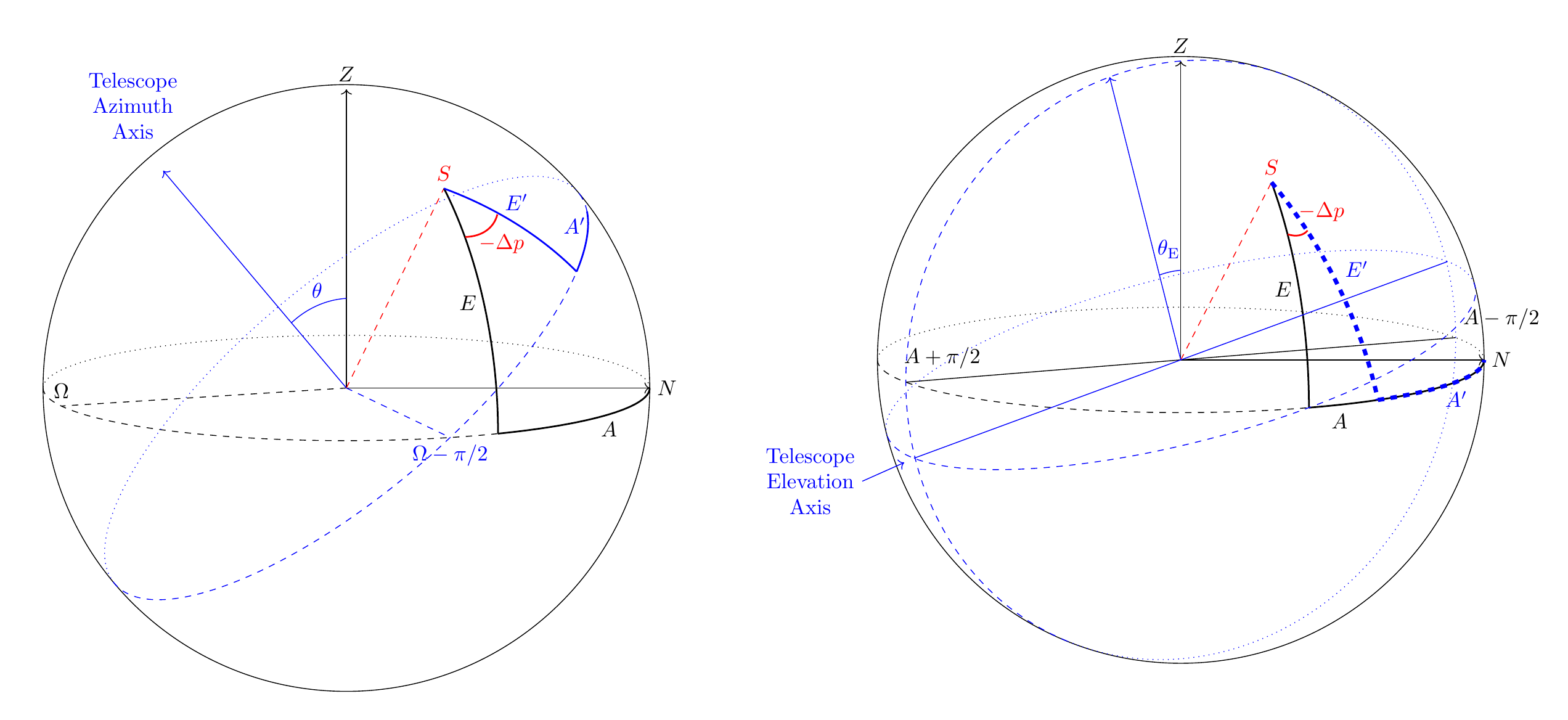}
\end{tabular}
\end{center}
\caption 
{ \label{fig:tilts}
A tilt of the azimuth (left) and elevation (right) axes of the telescope can cause a significant change in the apparent parallactic angle ($p$) of a target ($S$). The magnitude of the tilt in azimuth ($\theta$) and elevation ($\theta_{\rm E}$) axes has been grossly exaggerated for the purpose of this diagram.} \end{figure} 

To construct a model of the tilt of the azimuth and elevation axes of the Gemini South telescope we assumed that the instrument rotator was only compensating for the change in parallactic angle induced by these tilts. We collected measurements of the telescope elevation and azimuth and instrument rotator position on the 207 nights where GPI observations were taken with the rotator drive enabled. As the header stores the mechanical position of the telescope, we inverted the previous equations to compute the topocentric elevation and azimuth. Using these, we predicted the change in parallactic angle, and thus the position that the instrument rotator would need to be at to compensate for non-perpendicularity, for a given set of tilt parameters ($\theta$, $\Omega$, $\theta_E$). We performed a least squares minimization to determine the set of tilt parameters that best reproduce the instrument rotator position for ten roughly six-month periods over the last five years. The break points were chosen arbitrarily to be at the start and mid-point of each year except for years in which a major earthquake occurred near Cerro Pachon (2015 September 17 and 2019 January 19), and when a break point coincided with a period in which GPI was being used.

\begin{figure}
\begin{center}
\begin{tabular}{c}
\includegraphics[width=10cm]{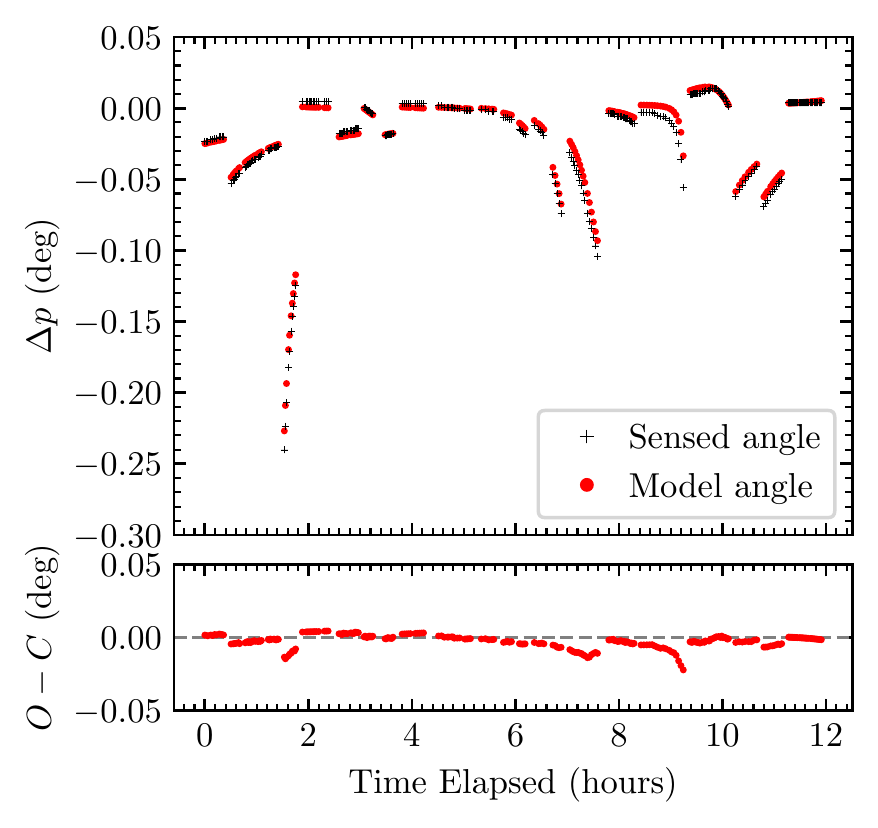}
\end{tabular}
\end{center}
\caption 
{ \label{fig:model_example}
Comparison between the sensed rotator angle (black) and that predicted by our simple telescope model (red) for observations taken on 2015 May 06 (top panel) and the corresponding residuals (bottom panel). The model is able to reproduce the sensed angle well for targets at low elevation (small values of $\Delta p$), but performs worse at very high elevations.} 
\end{figure} 

\begin{figure}
\begin{center}
\begin{tabular}{c}
\includegraphics[width=10cm]{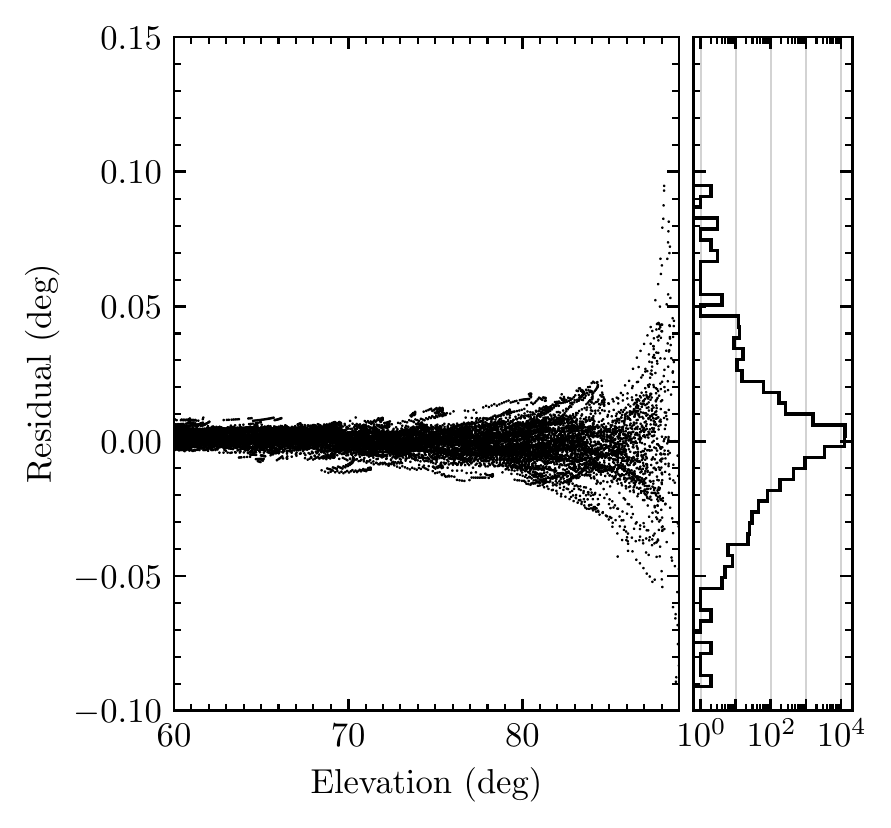}
\end{tabular}
\end{center}
\caption 
{ \label{fig:model_residual}
Residuals between the sensed rotator angle and that predicted by the model for all observations in the GPIES database where the rotator drive was enabled, plotted as a function of elevation (left panel), and as a marginalized histogram on a logarithmic scale (right panel). The residuals are significant for observations taken at an elevation of $E>88$\,deg; 33 of the 32,644 images in the database have a residual greater than 0.05 deg.} 
\end{figure} 

\begin{table}[ht]
\caption{Tilt model fit parameters} 
\label{tbl:tilt}
\begin{center}       
\begin{tabular}{cccccc}
Start Date & End Date & $\theta$ & $\Omega$ & $\theta_E$ & $N_{\rm frames}$\\
(UT)&(UT)&(arc sec)&(deg)&(arc sec)&\\
\hline
$\cdots$   & 2014-06-30 & 27.3 & 38.9 & 16.8 & 3406\\
2014-07-01 & 2014-12-31 & 27.3 & 42.8 & 15.4 & 2787\\
2015-01-01 & 2015-09-16 & 29.5 & 43.8 & 20.1 & 5013\\
2015-09-17 & 2016-06-30 & 29.2 & 45.5 & 16.8 & 6323\\
2016-07-01 & 2016-12-31 & 28.1 & 40.5 & 16.5 & 1806\\
2017-01-01 & 2017-07-02 & 27.5 & 50.2 & 18.1 & 1641\\
2017-07-03 & 2017-12-31 & 26.4 & 37.6 & 19.7 & 2751\\
2018-01-01 & 2018-06-30 & 29.0 & 44.5 & 15.2 & 4682\\
2018-07-01 & 2019-01-18 & 29.5 & 45.4 & 16.1 & 3001\\
2019-01-19 &   $\cdots$ & 31.1 & 51.8 & 19.1 & 1234
\end{tabular}
\end{center}
\end{table} 

The tilt model parameters that best fit the measured instrument rotator positions are given in Table~\ref{tbl:tilt}. A comparison between the model and data on the night of 2015 May 6 UT is shown in Figure~\ref{fig:model_example}. The model is able to reproduce the commanded rotator positions with residuals smaller than the north calibration uncertainty (discussed below) in all but a handful of the images, specifically those taken at elevations $\gtrsim 88$\,deg (Fig.~\ref{fig:model_residual}). 

We identified all GPI images to which we had access that were taken with the instrument rotator drive disabled. We used the tilt model parameters in Table~\ref{tbl:tilt} and the telescope elevation and azimuth within the header to calculate the correction to apply to the parallactic angle to compensate for the slight change in the angle of vertical on the detector. We created a lookup table with these corrections using the {\tt DATALAB} header keyword to uniquely assign a correction to a specific GPI observation taken with the rotator drive disabled. Files with {\tt DATALAB} values not in the lookup table do not have a correction applied. This lookup table contains all GPI observations taken with the drive disabled that were accessible at the time of this study, including GPIES campaign data, GO program data that are ingested into the GPIES database, and GO program data that was public at the time of the analysis.

\section{North Angle Calibration}
\label{sec:new_calib}
The corrections to the GPI DRP described in Sections~\ref{sec:updates}, \ref{sec:clocks}, and \ref{sec:rotator} necessitated a revision of GPI's astrometric calibration, specifically the true north angle. The north angle offset is defined as the angle between IFS pixel columns and North in an image that has been rotated to put North up based on the average parallactic angle during the exposure. Here we define the direction of the north angle offset as $\theta_{\rm true} - \theta_{\rm observed}$, a correction that would need to be added to a position angle measured in images reduced with the GPI DRP (after correcting for the $x$-axis flip) to recover the true position angle of a companion. 

We calibrate true north in GPI data based on observations of astrometric reference targets on sky. The small field of view ($2\farcs8 \times 2\farcs8$) and relatively bright limiting magnitude ($I<10$) of GPI exclude many of the typical astrometric calibration fields used by other instruments (e.g., M15, M92). Instead, we rely on periodic observations of a set of calibration binaries that have near-contemporaneous measurements with the well-calibrated NIRC2 camera on the Keck II telescope\cite{Yelda:2010ig,Service:2016gk}.

\subsection{Gemini South/GPI Observations}
\singlespacing
\begin{longtable}{cccccccccc}
\caption{GPI observing log} \\
\label{tbl:gpi_obs}
Target & UT Date  & Mode\footnote{C: coronagraphic, D: direct, ND: neutral density, U: unblocked} & Filter & $t_{\rm int}$ & $n_{\rm coadd}$ & $n_{\rm exp}$ & $\rho$ & $\theta$\\
&&&&(sec)&&&(px)&(deg)\\
\hline
HD 1620 & 2015-08-30 & C & H & 1.45 & 10 & 23 & $41.342\pm0.027$ & $181.529\pm0.040$ \\
HD 1620 & 2015-11-05 & D & H & 1.45 & 10 & 14 & $41.289\pm0.055$ & $181.202\pm0.069$ \\
HD 1620 & 2018-07-21 & U & H & 1.45 & 10 & 10 & $41.024\pm0.043$ & $178.457\pm0.058$ \\
HD 1620 & 2018-08-09 & U & H & 1.45 & 10 & 24 & $41.001\pm0.020$ & $178.434\pm0.037$ \\
HD 1620 & 2018-09-21 & U & H & 1.45 & 10 & 24 & $40.981\pm0.023$ & $178.334\pm0.039$ \\
HD 1620 & 2018-11-18 & U & H & 1.45 & 10 & 25 & $40.983\pm0.027$ & $178.272\pm0.069$ \\
HD 1620 & 2018-12-20 & U & H & 1.45 & 10 & 17 & $40.982\pm0.019$ & $178.058\pm0.068$ \\
HD 1620 & 2019-08-10 & U & H & 1.45 & 1 & 9 & $40.900\pm0.048$ & $(177.400\pm0.086)$ \\
HD 1620 & 2019-08-10 & U & H & 1.45 & 10 & 7 & $40.878\pm0.019$ & $(177.377\pm0.038)$ \\
HD 1620 & 2019-08-10 & U & H & 1.45 & 1, 10 & 16 & $40.882\pm0.039$ & $177.389\pm0.070$\footnote{Calculated using all images obtained on 2019-08-10} \\
\hline
HD 6307 & 2015-09-01 & D & H & 1.45 & 10 & 19 & $59.915\pm0.029$ & $237.081\pm0.066$ \\
HD 6307 & 2019-08-10 & U & H & 4.36 & 1 & 19 & $60.365\pm0.035$ & $236.659\pm0.032$ \\
\hline
HD 157516 & 2015-07-01 & D & H & 1.45 & 10 & 13 & $48.773\pm0.025$ & $142.511\pm0.019$ \\
HD 157516 & 2015-07-29 & D & H & 1.45 & 10 & 7 & $48.759\pm0.058$ & $142.514\pm0.052$ \\
HD 157516 & 2015-07-30 & D & H & 1.45 & 10 & 20 & $48.788\pm0.041$ & $142.457\pm0.027$ \\
\hline
HD 158614 & 2019-08-11 & ND & H & 5.82 & 5 & 14 & $26.238\pm0.017$ & $127.714\pm0.046$ \\
\hline
HIP 43947 & 2014-05-14 & D & H & 1.45 & 5 & 9 & $29.994\pm0.010$ & $(260.908\pm0.020)$ \\
HIP 43947 & 2014-05-14 & D & H & 1.45 & 1 & 20 & $29.994\pm0.056$ & $(260.909\pm0.059)$ \\
HIP 43947 & 2014-05-14 & D & H & 1.45 & 1, 5 & 29 & $29.994\pm0.047$ & $260.908\pm0.051$\footnote{Calculated using all images obtained on 2014-05-14} \\
HIP 43947 & 2015-01-24 & D & H & 1.45 & 5 & 12 & $29.991\pm0.015$ & $260.872\pm0.013$ \\
HIP 43947 & 2015-04-02 & D & H & 1.45 & 5 & 12 & $29.982\pm0.012$ & $260.878\pm0.016$ \\
HIP 43947 & 2015-04-23 & D & H & 1.45 & 5 & 12 & $29.978\pm0.016$ & $261.036\pm0.027$ \\
\hline
HIP 44804 & 2014-03-23 & D & K1 & 1.45 & 10 & 4 & $32.159\pm0.009$ & $306.035\pm0.027$ \\
HIP 44804 & 2014-05-14 & D & H & 1.45 & 5 & 14 & $32.096\pm0.069$ & $305.866\pm0.076$ \\
\hline
HIP 80628 & 2019-04-27 & ND & H & 8.73 & 3 & 12 & $69.367\pm0.014$ & $55.733\pm0.016$ \\
HIP 80628 & 2019-08-10 & ND & H & 8.73 & 3 & 9 & $69.770\pm0.012$ & $56.362\pm0.015$ \\
\hline
HR 7668 & 2016-09-21 & U & H & 1.45 & 10 & 5 & $37.337\pm0.012$ & $114.342\pm0.017$ \\
HR 7668 & 2016-09-21 & U & K1 & 1.45 & 10 & 15 & $37.332\pm0.021$ & $(114.308\pm0.052)$\footnote{These data are not used for deriving the plate scale and north angle in Section~\ref{sec:new_tn}} \\
\hline
$\theta^1$ Ori B2-B3 & 2014-09-12 & C & H & 29.10 & 1 & 12 & $8.143\pm0.059$ & $222.881\pm0.440$ \\
$\theta^1$ Ori B2-B3 & 2014-11-11 & C & H & 14.55 & 2 & 13 & $8.124\pm0.021$ & $223.718\pm0.165$ \\
$\theta^1$ Ori B2-B3 & 2014-12-17 & C & H & 29.10 & 1 & 10 & $8.067\pm0.028$ & $224.067\pm0.166$ \\
$\theta^1$ Ori B2-B3 & 2015-01-31 & C & H & 29.10 & 1 & 10 & $8.107\pm0.016$ & $223.816\pm0.103$ \\
$\theta^1$ Ori B2-B3 & 2015-04-06 & C & H & 29.10 & 1 & 8 & $8.085\pm0.031$ & $223.923\pm0.208$ \\
$\theta^1$ Ori B2-B3 & 2015-12-01 & C & H & 29.10 & 1 & 10 & $8.106\pm0.035$ & $224.920\pm0.196$ \\
$\theta^1$ Ori B2-B3 & 2015-12-19 & C & H & 29.10 & 1 & 10 & $8.114\pm0.024$ & $224.894\pm0.156$ \\
$\theta^1$ Ori B2-B3 & 2016-01-21 & C & H & 29.10 & 1 & 10 & $8.084\pm0.020$ & $225.076\pm0.135$ \\
$\theta^1$ Ori B2-B3 & 2016-02-26 & C & H & 8.73 & 1 & 15 & $8.113\pm0.037$ & $224.837\pm0.330$ \\
$\theta^1$ Ori B2-B3 & 2016-03-18 & C & H & 8.73 & 3 & 7 & $8.088\pm0.017$ & $225.059\pm0.122$ \\
$\theta^1$ Ori B2-B3 & 2016-09-19 & C & H & 29.10 & 1 & 10 & $8.102\pm0.020$ & $225.845\pm0.146$ \\
$\theta^1$ Ori B2-B3 & 2016-11-17 & C & H & 29.10 & 1 & 10 & $8.073\pm0.026$ & $226.106\pm0.162$ \\
$\theta^1$ Ori B2-B3 & 2016-12-21 & C & H & 8.73 & 3 & 9 & $8.096\pm0.023$ & $225.936\pm0.135$ \\
$\theta^1$ Ori B2-B3 & 2017-02-13 & C & H & 8.73 & 3 & 10 & $8.070\pm0.018$ & $226.289\pm0.141$ \\
$\theta^1$ Ori B2-B3 & 2017-04-20 & C & H & 8.73 & 3 & 10 & $8.072\pm0.027$ & $226.430\pm0.158$ \\
$\theta^1$ Ori B2-B3 & 2017-11-06 & C & H & 8.73 & 3 & 10 & $8.087\pm0.026$ & $226.947\pm0.176$ \\
$\theta^1$ Ori B2-B3 & 2017-11-10 & C & H & 8.73 & 6 & 3 & $8.076\pm0.019$ & $226.860\pm0.089$ \\
$\theta^1$ Ori B2-B3 & 2018-01-06 & C & H & 8.73 & 6 & 7 & $8.085\pm0.010$ & $227.008\pm0.074$ \\
$\theta^1$ Ori B2-B3 & 2018-01-29 & C & H & 8.73 & 6 & 7 & $8.062\pm0.028$ & $227.410\pm0.262$ \\
$\theta^1$ Ori B2-B3 & 2018-03-08 & C & H & 8.73 & 6 & 7 & $8.078\pm0.029$ & $227.127\pm0.195$ \\
$\theta^1$ Ori B2-B3 & 2018-03-24 & C & H & 24.73 & 2 & 11 & $8.083\pm0.032$ & $227.303\pm0.177$ \\
$\theta^1$ Ori B2-B3 & 2018-03-26 & C & H & 14.55 & 4 & 7 & $8.071\pm0.011$ & $227.481\pm0.090$ \\
$\theta^1$ Ori B2-B3 & 2018-04-07 & C & H & 8.73 & 6 & 2 & $8.127\pm0.060$ & $227.125\pm0.366$ \\
$\theta^1$ Ori B2-B3 & 2018-11-19 & C & H & 14.55 & 4 & 7 & $8.091\pm0.016$ & $228.055\pm0.072$ \\
$\theta^1$ Ori B2-B3 & 2019-08-10 & C & H & 14.55 & 4 & 6 & $8.070\pm0.032$ & $228.762\pm0.182$ \\
\hline
\end{longtable} 
\doublespacing

We have observed nine binary or multiple star systems since the start of routine operations in 2014.  A summary of all these observations is given in Table~\ref{tbl:gpi_obs}.  These observations were obtained with GPI's $H$ band filter ($\lambda_{\rm eff}=1.64$\,$\mu$m) for all except two sequences taken with the $K1$ filter ($\lambda_{\rm eff}=2.06$\,$\mu$m); note that since the spectral filter in the GPI IFS is after the spatial pixellation at the lenslet array, change of filter cannot affect the astrometric calibration. The majority of the observations were obtained in GPI's ``direct'' mode, a configuration where the various coronagraphic components are removed from the optical path. Some were obtained in ``unblocked'' mode, which includes the Lyot mask and pupil plane apodizer in the optical path to reduce instrument throughput, preventing saturation on brighter stars. The addition of a neutral density filter in 2017 allowed us to observe calibrator binaries that were significantly brighter than the nominal $H$-band saturation limit of the IFS in either ``direct'' or ``unblocked'' mode. Observations of the $\theta^1$ Ori B multiple system were taken in the coronagraphic mode, the typical mode for planet search observations, allowing for a high signal-to-noise detection of the fainter stellar components B2, B3 and B4 that all lie within an arcsecond of the primary star.

We do not expect the coronagraph optics to have a significant effect on astrometric measurements, except for those made for objects extremely close to the edge of the focal plane mask which is not relevant here. The three coronagraph optics are in pupil and focal planes only, so cannot individually introduce distortions. By effectively weighting the beam profile across the pupil, they could in principle cause the beam to sample a different portion of any intermediate optics; if those optics have polishing errors that could cause a slight field-dependent photocenter shift. However, this effect should be neglible. The intermediate optics (see Fig.~\ref{fig:gpi_diagram}) are small, located in a slow beam, and superpolished to $\sim$1\,nm RMS wavefront error. Measured distortions are $\sim$3\,mas across the field of view\cite{Konopacky:2014hf} and completely dominated by the geometric effects of the telephoto relay inside the spectrograph, with no evidence for a polishing-error component.

These observations were processed using version 1.5 (revision {\tt e0ea9f5}) of the GPI DRP, incorporating the changes described in Sections~\ref{sec:updates}, \ref{sec:clocks}, and \ref{sec:rotator}. The data were all processed using the same DRP recipe with standard processing steps. The raw images were dark subtracted, and corrected for bad pixels using both a static bad pixel map and outlier identification. The individual microspectra in each two-dimensional image were reassembled into a three-dimensional data cube ($x$,$y$,$\lambda$) using a wavelength solution derived from  observations of calibration argon arc lamp. An additional outlier identification and rejection step was performed on the individual slices of the data cubes. A distortion correction was then applied to each slice based on measurements of a pinhole mask taken during the commissioning of the instrument \cite{Konopacky:2014hf}.

\subsection{Keck II/NIRC2 Observations}
\label{sec:nirc2}
\singlespacing
\begin{longtable}{ccccccccc}
\caption{NIRC2 observing log} \\
\label{tbl:nirc2_obs}
Target & UT Date & Filter & Rot. &  $t_{\rm int}$ & $n_{\rm coadd}$ & $n_{\rm exp}$ & $\rho$ & $\theta$\\
&&&Mode&(sec)&&&(mas)&(deg)\\
HD 1620 & 2015-08-02 & H2$_{2-1}$ & PA & 0.18 & 50 & 9 & $585.93\pm0.41$ & $181.740\pm0.030$ \\
\hline
HD 6307 & 2015-08-02 & $K^{\prime}$ & PA & 0.181 & 100 & 9 & $848.36\pm0.48$ & $237.136\pm0.031$ \\
\hline
HD 157516 & 2015-05-11 & $K_{\rm cont}$ & VA & 1.0 & 15 & 3 & $690.57\pm0.50$ & $142.678\pm0.035$ \\
\hline
HD 158614 & 2014-05-13 & Br$\gamma$ & PA & 0.053 & 100 & 12 & $787.50\pm0.35$ & $147.406\pm0.017$ \\
HD 158614 & 2019-08-17 & Br$\gamma$ & PA & 0.053 & 100 & 45 & $369.64\pm0.22$ & $128.014\pm0.026$ \\
HD 158614 & 2019-08-26 & Br$\gamma$ & PA & 0.053 & 100 & 42 & $367.36\pm0.22$ & $127.799\pm0.023$ \\
\hline
HIP 43947 & 2014-03-13 & $K^{\prime}$ & VA & 1.0 & 1 & 4 & $424.70\pm0.46$ & $260.948\pm0.039$ \\
\hline
HIP 44804 & 2014-03-13 & Br$\gamma$ & VA & 0.5 & 5 & 4 & $455.19\pm0.68$ & $306.325\pm0.021$ \\
HIP 44804 & 2019-05-23 & $H_{\rm cont}$ & VA & 2.0 & 10 & 16 & $444.73\pm0.31$ & $297.145\pm0.049$ \\
\hline
HIP 80628 & 2014-03-13 & Br$\gamma$ & VA & 0.181 & 1 & 4 & $859.09\pm0.39$ & $43.640\pm0.024$ \\
HIP 80628 & 2019-04-25 & $H_{\rm cont}$ & PA & 0.1 & 50 & 9 & $982.01\pm0.54$ & $56.187\pm0.025$ \\
HIP 80628 & 2019-05-15 & $H_{\rm cont}$ & PA & 0.017 & 100 & 14 & $983.13\pm0.59$ & $56.327\pm0.026$ \\
HIP 80628 & 2019-05-23 & $H_{\rm cont}$ & VA & 0.01 & 100 & 10 & $983.64\pm0.63$ & $56.242\pm0.023$ \\
HIP 80628 & 2019-08-17 & Br$\gamma$ & PA & 0.0176 & 100 & 33 & $988.58\pm0.65$ & $56.855\pm0.027$ \\
HIP 80628 & 2019-08-26 & Br$\gamma$ & PA & 0.0176 & 100 & 42 & $989.20\pm0.56$ & $56.881\pm0.024$ \\
HIP 80628 & 2019-08-26 & Br$\gamma$ & VA & 0.0176 & 100 & 42 & $988.90\pm0.63$ & $56.863\pm0.024$ \\
\hline
HR 7668 & 2016-07-22 & Br$\gamma$ & PA & 1.0 & 10 & 9 & $528.55\pm0.41$ & $114.725\pm0.034$ \\
\hline
$\theta^1$ Ori B2-B3 & 2001-12-20 & NB$_{\rm 2.108}$ & PA & 0.2 & 25 & 6 & $115.69\pm0.40$ & $209.32\pm0.20$ \\
$\theta^1$ Ori B2-B3 & 2004-10-03 & Br$\gamma$ & PA & 0.2 & 100 & 2 & $116.97\pm0.77$ & $212.17\pm0.38$ \\
$\theta^1$ Ori B2-B3 & 2005-02-16 & NB$_{\rm 2.108}$ & PA & 0.2 & 50 & 3 & $116.34\pm0.45$ & $212.70\pm0.22$ \\
$\theta^1$ Ori B2-B3 & 2005-02-25 & Br$\gamma$ & PA & 0.2 & 50 & 3 & $116.93\pm0.30$ & $212.94\pm0.15$ \\
$\theta^1$ Ori B2-B3 & 2011-02-06 & Br$\gamma$ & VA & 0.726 & 1 & 6 & $114.97\pm0.89$ & $219.47\pm0.44$ \\
$\theta^1$ Ori B2-B3 & 2011-02-06 & Br$\gamma$ & PA & 0.726 & 1 & 9 & $116.03\pm0.71$ & $219.35\pm0.35$ \\
$\theta^1$ Ori B2-B3 & 2014-09-03 & $K^{\prime}$ & VA & 0.032 & 100 & 6 & $115.12\pm0.14$ & $223.90\pm0.07$ \\
$\theta^1$ Ori B2-B3 & 2014-12-06 & $H$ & VA & 0.053 & 100 & 15 & $115.41\pm0.28$ & $223.99\pm0.14$ \\
$\theta^1$ Ori B2-B3 & 2015-10-27 & Br$\gamma$ & PA & 0.75 & 30 & 9 & $115.07\pm0.23$ & $224.93\pm0.11$ \\
$\theta^1$ Ori B2-B3 & 2016-01-18 & Br$\gamma$ & PA & 0.75 & 30 & 10 & $115.52\pm0.20$ & $225.08\pm0.10$ \\
$\theta^1$ Ori B2-B3 & 2016-02-04 & Br$\gamma$ & VA & 0.75 & 30 & 6 & $114.88\pm0.16$ & $225.14\pm0.08$ \\
$\theta^1$ Ori B2-B3 & 2016-02-21 & Br$\gamma$ & PA & 0.181 & 300 & 9 & $115.17\pm0.19$ & $225.13\pm0.09$ \\
$\theta^1$ Ori B2-B3 & 2016-08-20 & $K_s$ & PA & 0.181 & 1 & 4 & $115.52\pm0.44$ & $226.23\pm0.22$ \\
$\theta^1$ Ori B2-B3 & 2018-02-13 & Br$\gamma$ & PA & 0.75 & 1 & 11 & $115.31\pm0.18$ & $227.59\pm0.08$ \\
\hline
\end{longtable} 
\doublespacing
The same nine multiple systems have been observed with the NIRC2 instrument in conjunction with the facility adaptive optics system on the Keck II telescope. The isolated calibration binaries have between one and six NIRC2 epochs between 2014 and 2019. The Trapezium cluster that contains $\theta^1$ Ori B has been observed periodically with NIRC2 as an astrometric calibrator field by multiple different teams, with archival measurements extending as far back as December 2001. The observations were taken in a variety of instrument configurations and filters. A summary of these observations is given in Table~\ref{tbl:nirc2_obs}. Datasets were taken in either position angle (PA) mode, where North remains fixed at a given angle on the detector, or vertical angle (VA) mode, where the vertical angle remains fixed and North varies with the parallactic angle of the target.

We reduced these data using a typical near-infrared imaging data reduction pipeline; correction for non-linearity \cite{Metchev:2009ky}, dark subtraction, flat fielding, and bad pixel identification and correction. Reduced images were corrected for geometric distortion using the appropriate distortion map\cite{Yelda:2010ig,Service:2016gk}. For observations taken using a subarray of the NIRC2 detector, we zero-padded the images prior to applying the distortion correction as the distortion correction script is hard-coded for $1024\times1024$\,px images\footnote{\url{https://github.com/jluastro/nirc2_distortion}}. The astrometric calibration of NIRC2 was derived from analyses of globular cluster observations, and has been validated with measurements of the locations of SiO masers in the galactic center that were determined precisely using very long baseline radio interferometry measurements\cite{Yelda:2010ig,Service:2016gk}. We used a plate scale of $9.952\pm0.002$\,mas\,px$^{-1}$ and a north angle offset of $-0.252\pm0.009$\,deg for data taken prior to 2015 April 13\cite{Yelda:2010ig}, and $9.971\pm0.005$\,mas\,px$^{-1}$ and a north angle offset of $-0.262\pm0.020$\,deg for data taken after\cite{Service:2016gk}.

\subsection{Relative Astrometry}
We used PSF fitting to measure the position of the companion relative to the primary. For the calibration binaries other than $\theta^1$ Ori B, we estimated the location of the primary star within each image (or wavelength slice) by fitting a two-dimensional Gaussian to a small $7\times7$\,pixel stamp centered on an initial estimate of the primary star. The five parameters ($x$, $y$, $\sigma_x$, $\sigma_y$ and amplitude $A$) were allowed to vary except for the NIRC2 data obtained on 2019-04-25 (HIP 80628) and 2019-05-23 (HIP 44804) where $\sigma_x$ and $\sigma_y$ were fixed due to a strongly asymmetric PSF and the proximity of the companion. This process was repeated using the output of the first iteration as the initial guess for the second. We extracted a $15\times15$\,px stamp centered on the fitted position of the primary to use as a template to fit the location of the secondary. We used the Nelder-Mead downhill simplex algorithm to determine the pixel offset and flux ratio between the primary and secondary star by minimizing the squared residuals within a $2\lambda/D$ radius aperture surrounding the secondary. We estimated the uncertainty in the centroid of each fit as the full-width-at-half-maximum divided by the signal to noise ratio measured as the peak pixel value divided by the standard deviation of pixel values within an annulus $15\lambda/D$ from the star. We corrected differential atmospheric refraction caused by the different zenith angle of the two stars using the model described in Ref.~\citenum{Gubler:1998fg}. We used the simplifying assumption that the observations were monochromatic at the central wavelength of the filter, negating any stellar color dependence on the effective wavelength. This effect causes a reduction in the separation of a binary star along the elevation axis, and was typically very small; at most 0.3\, mas for the NIRC2 observations of HIP 80628 taken at an elevation of $\sim$35\,deg. Position angles measured in datasets taken in VA mode were corrected by the parallactic angle at the middle of the exposure such that they were effectively measured relative from North.

The small angular separation between the two components of the $\theta^1$ Ori B2-B3 binary required us to use either $\theta^1$ Ori B1 for the NIRC2 observations or $\theta^1$ Ori B4 for the GPI observations as a reference PSF. We used this template PSF to simultaneously fit the location and fluxes of the two components of the B2-B3 binary following a similar procedure. We used a Fourier high-pass filter to subtract the seeing halo from B1 that was introducing a background signal for both B4 and the B2-B3 binary. The relative astrometry are listed in Table~\ref{tbl:gpi_obs} for GPI and in Table~\ref{tbl:nirc2_obs} for NIRC2. We did not apply any correction for the differential atmospheric refraction for these observations given the extremely small difference in zenith angle between the two stars. We did not use the relative astrometry of B1-B2, B1-B3, or B1-B4 as B1 was obscured by GPI's focal plane mask, nor did we use B2-B4 or B3-B4 as the relative motion of these three stars cannot be described using a simple Keplerian model.

As a verification of the relative astrometry presented here, we performed an independent analysis of a subset of both the GPI and NIRC2 observations using the procedure described in Ref.~\citenum{Konopacky:2014hf}. The GPI data were reduced with the same version of the DRP, while the NIRC2 data were reduced with a separate pipeline that performed the same functions as described in Section~\ref{sec:nirc2}. Once the data were reduced, relative astrometry was performed using {\tt StarFinder}\cite{Diolaiti:2000dt}. For this subset of observations we measured consistent separations and position angles to the values reported in Tables~\ref{tbl:gpi_obs} and \ref{tbl:nirc2_obs}.

\subsection{Accounting for Orbital Motion}
Orbital motion of the calibration binaries between the NIRC2 and GPI epochs can introduce a significant bias in the north angle offset measurement. We fit Keplerian orbits to each of the calibration binaries using the NIRC2 astrometry presented in Table~\ref{tbl:nirc2_obs}. These fits allowed us to simulate NIRC2 measurements on the same epoch as the GPI observations listed in Table~\ref{tbl:gpi_obs}, mitigating the bias induced by orbital motion. We use the parallel-tempered affine invariant Markov chain Monte Carlo (MCMC) package {\tt emcee}\cite{ForemanMackey:2013io} to sample the posterior distributions of the Campbell elements describing the visual orbit and of the system parallax. A complete description of the fitting procedure as applied to the 51 Eridani system can be found in reference \citenum{DeRosa:2020gy}. We used prior distributions for the system mass based on the blended spectral type and flux ratios of the components, and for the system parallax using measurements from either {\it Hipparcos}\cite{vanLeeuwen:2007dc} or {\it Gaia}\cite{GaiaCollaboration:2018io}. We used a parallax of $2.41\pm0.03$\,mas for $\theta^1$ Ori B2-B3\cite{Reid:2014km}. We also fitted the radial velocity measurements of both components of the HD 158614 binary\cite{Pourbaix:2004dg} to help further constrain its orbital parameters. We purposely excluded astrometric measurements from other instruments and assumed that the NIRC2 astrometric calibration was stable before and after the realignment procedure in mid-2015.

\begin{figure}
\begin{center}
\begin{tabular}{c}
\includegraphics[width=16cm]{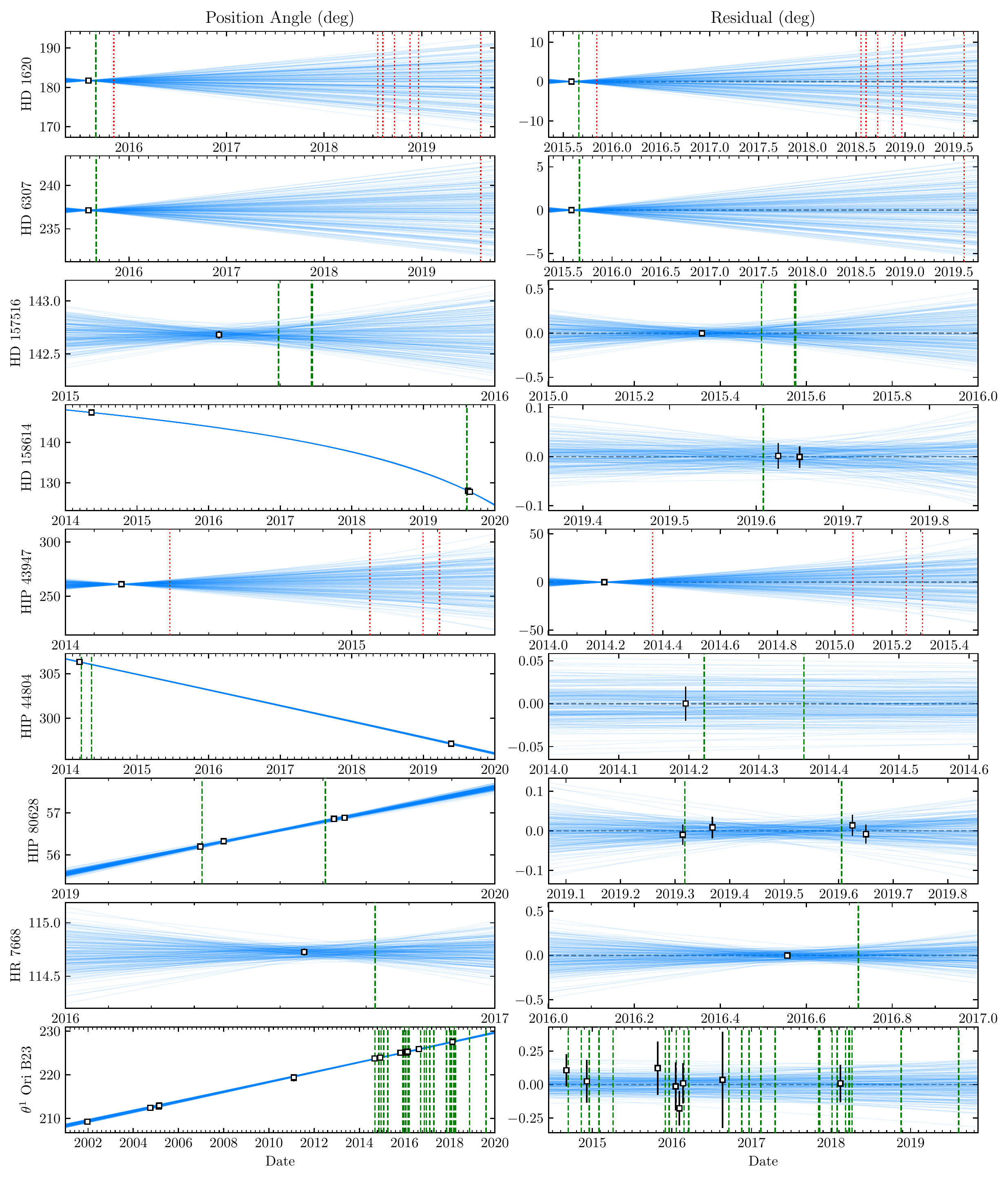}
\end{tabular}
\end{center}
\caption 
{ \label{fig:orbital_motion}
Position angle (left column) and residuals (right column) of the orbits (blue lines) consistent with the NIRC2 astrometry in Table~\ref{tbl:nirc2_obs} (squares). The dates of GPI observations are highlighted; green dashed lines denote epochs that were used for the astrometric calibration, and red dotted lines denote epochs where the orbital motion is significant relative to the GPI measurement uncertainties. In a subset of the plots in the right column the date range has been restricted to focus on the dates of the GPI observations.} 
\end{figure} 

The position angle of the visual orbit and correspond residuals are shown in Figure~\ref{fig:orbital_motion} for the nine calibration binaries. We simulated NIRC2 measurements at the epoch of the GPI observations by drawing 10,000 orbits at random from MCMC chains and converting the orbital elements into separations and position angles at the desired epoch. We used the median of the resulting distribution of separations and position angles as the simulated measurement and the standard deviation as the uncertainty. These simulated measurements are reported in Table~\ref{tbl:tn_log}. The small semi-major axis of the HIP 43947 binary led to a significant uncertainty on the simulated NIRC2 observation despite the short fifty-day baseline between the NIRC2 and GPI observations, precluding a measurement of the north offset angle with this binary. This was also the case for all but one epoch of both the HD 1620 and HD 6307 systems. Additional observations of these systems with NIRC2 to reduce the orbital uncertainties will be required for more precise predictions at these epochs. The remaining binaries (HD 157516, HD 158614, HIP 44804, HIP 80628, HR 7668, and $\theta^1$ Ori B2-B3) either had enough NIRC2 measurements to sufficiently constrain the orbit at the GPI epochs, or were close enough in time that the orbital motion between the NIRC2 and GPI epochs was smaller than the measurement uncertainties.

\section{Revised Astrometric Calibration}
\label{sec:new_tn}
\subsection{GPI Plate Scale}
\begin{figure}
\begin{center}
\begin{tabular}{c}
\includegraphics[width=10cm]{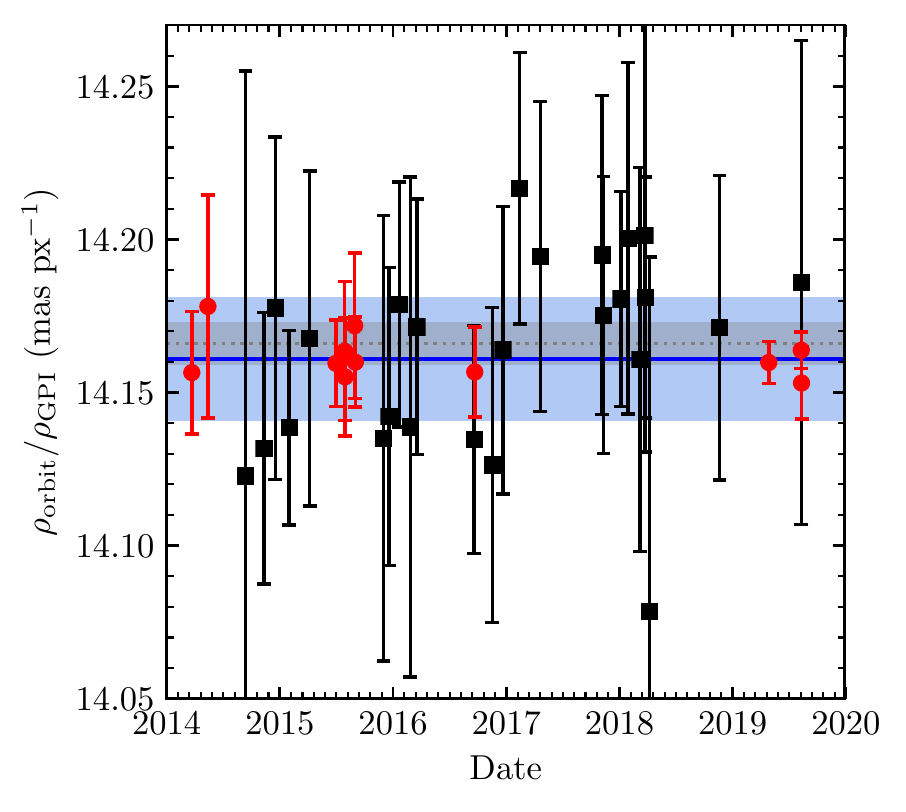}
\end{tabular}
\end{center}
\caption 
{ \label{fig:ps}
Measurements of the plate scale of GPI derived from calibration binaries (red circles) and the $\theta^1$ Ori B2-B3 binary (black squares).  The mean and standard deviation (blue solid line and shaded region) were calculated using a weighted mean and assuming that the measurements were not independent. The previous astrometric calibration is overplotted for reference (grey dashed line and shaded region).} 
\end{figure}

The plate scale for GPI was measured using the predicted separations in angular units from the orbit fit to the NIRC2 measurements and the pixel separations measured in the reduced GPI images (Table~\ref{tbl:tn_log}). We saw no evidence of a variation in the plate scale with time (Fig.~\ref{fig:ps}), and adopted a single value of $14.161\pm0.021$\,mas\,px$^{-1}$. This measurement is consistent with the previous plate scale of $14.166\pm0.007$\,mas\,px$^{-1}$\cite{DeRosa:2015jl,Konopacky:2014hf}, but with a larger uncertainty. The pipeline changes described in Sections~\ref{sec:updates}, \ref{sec:clocks}, and \ref{sec:rotator} have no impact on the separation of two stars within a reduced GPI image.  The slight difference in the inferred plate scale can instead be ascribed to changes in the way the relative positions of the two components of each calibration binary were measured, the greater number of measurements, or simply to measurement uncertainties.

\subsection{GPI North Offset Angle}
\singlespacing
\begin{longtable}{cccccc}
\caption{GPI plate scale and north offset angle} \\
\label{tbl:tn_log}
UT Date & Target & $\rho_{\rm orbit}$ & $\theta_{\rm orbit}$ & $\rho_{\rm orbit}/\rho_{\rm GPI}$ & $\theta_{\rm orbit}-\theta_{\rm GPI}$\\
&&(mas)&(deg)&(mas\,px$^{-1}$)&(deg)\\
\hline
2014-03-23 & HIP 44804 & $455.26\pm0.63$ & $306.274\pm0.020$ & $14.157\pm0.020$ & $0.239\pm0.034$ \\
2014-05-14 & HIP 43947 & $424.81\pm12.42$ & $260.872\pm2.154$ & $(14.163\pm0.415)$ & $(-0.036\pm2.155)$ \\
2014-05-14 & HIP 44804 & $455.06\pm0.64$ & $306.028\pm0.020$ & $14.178\pm0.036$ & $0.162\pm0.079$ \\
\multicolumn{6}{r}{Weighted mean (2013-11-11 to 2014-09-08) : $0.23\pm0.11$\,deg} \\
\hline
2014-09-12 & $\theta^1$ Ori B2-B3 & $114.88\pm0.14$ & $223.640\pm0.052$ & $14.123\pm0.132$ & $0.239\pm0.414$ \\
2014-11-11 & $\theta^1$ Ori B2-B3 & $114.87\pm0.14$ & $223.826\pm0.052$ & $14.132\pm0.044$ & $0.189\pm0.198$ \\
2014-12-17 & $\theta^1$ Ori B2-B3 & $114.86\pm0.13$ & $223.938\pm0.052$ & $14.178\pm0.056$ & $-0.143\pm0.185$ \\
2015-01-24 & HIP 43947 & $418.33\pm65.15$ & $260.540\pm11.398$ & $(13.949\pm2.172)$ & $(-0.332\pm11.398)$ \\
2015-01-31 & $\theta^1$ Ori B2-B3 & $114.86\pm0.13$ & $224.078\pm0.052$ & $14.138\pm0.032$ & $0.241\pm0.124$ \\
2015-04-02 & HIP 43947 & $414.66\pm80.02$ & $260.435\pm14.097$ & $(13.830\pm2.669)$ & $(-0.443\pm14.097)$ \\
2015-04-06 & $\theta^1$ Ori B2-B3 & $114.85\pm0.12$ & $224.283\pm0.053$ & $14.168\pm0.055$ & $0.308\pm0.225$ \\
2015-04-23 & HIP 43947 & $413.34\pm84.75$ & $260.404\pm14.978$ & $(13.788\pm2.827)$ & $(-0.632\pm14.978)$ \\
2015-07-01 & HD 157516 & $690.60\pm0.59$ & $142.678\pm0.053$ & $14.159\pm0.014$ & $0.167\pm0.056$ \\
2015-07-29 & HD 157516 & $690.60\pm0.74$ & $142.678\pm0.071$ & $14.164\pm0.023$ & $0.164\pm0.088$ \\
2015-07-30 & HD 157516 & $690.60\pm0.74$ & $142.678\pm0.072$ & $14.155\pm0.019$ & $0.221\pm0.077$ \\
2015-08-30 & HD 1620 & $585.89\pm0.90$ & $181.740\pm0.105$ & $14.172\pm0.024$ & $0.211\pm0.112$ \\
2015-09-01 & HD 6307 & $848.39\pm0.78$ & $237.138\pm0.062$ & $14.160\pm0.015$ & $0.057\pm0.091$ \\
\multicolumn{6}{r}{Weighted mean (2014-09-08 to 2015-10-31) : $0.17\pm0.14$\,deg} \\
\hline
2015-11-05 & HD 1620 & $585.88\pm2.82$ & $181.727\pm0.346$ & $(14.190\pm0.071)$ & $(0.525\pm0.353)$ \\
2015-12-01 & $\theta^1$ Ori B2-B3 & $114.84\pm0.11$ & $225.024\pm0.055$ & $14.135\pm0.073$ & $0.052\pm0.200$ \\
2015-12-19 & $\theta^1$ Ori B2-B3 & $114.83\pm0.11$ & $225.080\pm0.055$ & $14.142\pm0.049$ & $0.204\pm0.214$ \\
2016-01-21 & $\theta^1$ Ori B2-B3 & $114.83\pm0.11$ & $225.183\pm0.055$ & $14.179\pm0.040$ & $0.197\pm0.177$ \\
2016-02-26 & $\theta^1$ Ori B2-B3 & $114.83\pm0.11$ & $225.295\pm0.056$ & $14.139\pm0.082$ & $0.319\pm0.491$ \\
2016-03-18 & $\theta^1$ Ori B2-B3 & $114.83\pm0.11$ & $225.360\pm0.056$ & $14.171\pm0.042$ & $0.321\pm0.155$ \\
\multicolumn{6}{r}{Weighted mean (2015-10-31 to 2016-09-05) : $0.21\pm0.23$\,deg} \\
\hline
2016-09-19 & $\theta^1$ Ori B2-B3 & $114.82\pm0.13$ & $225.938\pm0.059$ & $14.135\pm0.037$ & $0.166\pm0.144$ \\
2016-09-21 & HR 7668 & $528.57\pm0.52$ & $114.727\pm0.055$ & $14.157\pm0.015$ & $0.385\pm0.058$ \\
2016-11-17 & $\theta^1$ Ori B2-B3 & $114.82\pm0.14$ & $226.121\pm0.060$ & $14.126\pm0.051$ & $0.099\pm0.190$ \\
2016-12-21 & $\theta^1$ Ori B2-B3 & $114.81\pm0.15$ & $226.227\pm0.061$ & $14.164\pm0.047$ & $0.287\pm0.163$ \\
2017-02-13 & $\theta^1$ Ori B2-B3 & $114.81\pm0.16$ & $226.394\pm0.062$ & $14.217\pm0.044$ & $0.211\pm0.184$ \\
2017-04-20 & $\theta^1$ Ori B2-B3 & $114.81\pm0.17$ & $226.603\pm0.064$ & $14.194\pm0.051$ & $0.223\pm0.176$ \\
\multicolumn{6}{r}{Weighted mean (2016-09-05 to 2017-10-13) : $0.32\pm0.15$\,deg} \\
\hline
2017-11-06 & $\theta^1$ Ori B2-B3 & $114.81\pm0.22$ & $227.224\pm0.069$ & $14.195\pm0.052$ & $0.254\pm0.220$ \\
2017-11-10 & $\theta^1$ Ori B2-B3 & $114.81\pm0.22$ & $227.237\pm0.069$ & $14.175\pm0.045$ & $0.303\pm0.131$ \\
2018-01-06 & $\theta^1$ Ori B2-B3 & $114.81\pm0.23$ & $227.414\pm0.071$ & $14.181\pm0.035$ & $0.377\pm0.109$ \\
2018-01-29 & $\theta^1$ Ori B2-B3 & $114.81\pm0.24$ & $227.485\pm0.072$ & $14.200\pm0.057$ & $0.179\pm0.275$ \\
2018-03-08 & $\theta^1$ Ori B2-B3 & $114.81\pm0.25$ & $227.603\pm0.073$ & $14.161\pm0.063$ & $0.369\pm0.202$ \\
2018-03-24 & $\theta^1$ Ori B2-B3 & $114.81\pm0.26$ & $227.653\pm0.073$ & $14.201\pm0.071$ & $0.234\pm0.218$ \\
2018-03-26 & $\theta^1$ Ori B2-B3 & $114.81\pm0.26$ & $227.660\pm0.073$ & $14.181\pm0.039$ & $0.162\pm0.114$ \\
2018-04-07 & $\theta^1$ Ori B2-B3 & $114.81\pm0.26$ & $227.700\pm0.074$ & $14.078\pm0.116$ & $0.179\pm0.534$ \\
2018-07-21 & HD 1620 & $584.69\pm32.23$ & $181.531\pm3.961$ & $(14.252\pm0.786)$ & $(3.074\pm3.961)$ \\
2018-08-09 & HD 1620 & $584.64\pm32.79$ & $181.528\pm4.030$ & $(14.259\pm0.800)$ & $(3.094\pm4.030)$ \\
\multicolumn{6}{r}{Weighted mean (2017-10-13 to 2018-09-01) : $0.28\pm0.19$\,deg} \\
\hline
2018-09-21 & HD 1620 & $584.52\pm34.08$ & $181.520\pm4.189$ & $(14.263\pm0.832)$ & $(3.186\pm4.189)$ \\
2018-11-18 & HD 1620 & $584.41\pm35.81$ & $181.509\pm4.402$ & $(14.260\pm0.874)$ & $(3.237\pm4.403)$ \\
2018-11-19 & $\theta^1$ Ori B2-B3 & $114.81\pm0.33$ & $228.402\pm0.081$ & $14.171\pm0.050$ & $0.267\pm0.113$ \\
2018-12-20 & HD 1620 & $584.28\pm36.77$ & $181.504\pm4.520$ & $(14.257\pm0.897)$ & $(3.446\pm4.521)$ \\
2019-04-27 & HIP 80628 & $982.22\pm0.43$ & $56.215\pm0.020$ & $14.160\pm0.007$ & $0.482\pm0.026$ \\
2019-08-10 & $\theta^1$ Ori B2-B3 & $114.83\pm0.42$ & $229.224\pm0.091$ & $14.186\pm0.079$ & $0.299\pm0.220$ \\
2019-08-10 & HD 1620 & $583.63\pm43.77$ & $181.462\pm5.385$ & $(14.276\pm1.071)$ & $(4.073\pm5.385)$ \\
2019-08-10 & HD 6307 & $847.62\pm30.81$ & $237.156\pm2.673$ & $(14.042\pm0.510)$ & $(0.497\pm2.673)$ \\
2019-08-10 & HIP 80628 & $988.21\pm0.38$ & $56.801\pm0.016$ & $14.164\pm0.006$ & $0.439\pm0.022$ \\
2019-08-11 & HD 158614 & $371.35\pm0.19$ & $128.153\pm0.017$ & $14.153\pm0.012$ & $0.439\pm0.049$ \\
\multicolumn{6}{r}{Weighted mean (2018-09-01 to 2019-08-27) : $0.45\pm0.11$\,deg} \\
\hline
\hline
\multicolumn{6}{r}{Weighted mean (all): $0.36\pm0.12$\,deg}\\
\multicolumn{6}{r}{$14.161\pm0.021$\,mas\,px$^{-1}$}\\
\hline
\multicolumn{5}{l}{\footnotesize Note: measurements in parentheses are not included in weighted mean.}
\end{longtable} 
\doublespacing

\begin{figure}
\begin{center}
\begin{tabular}{c}
\includegraphics[width=16cm]{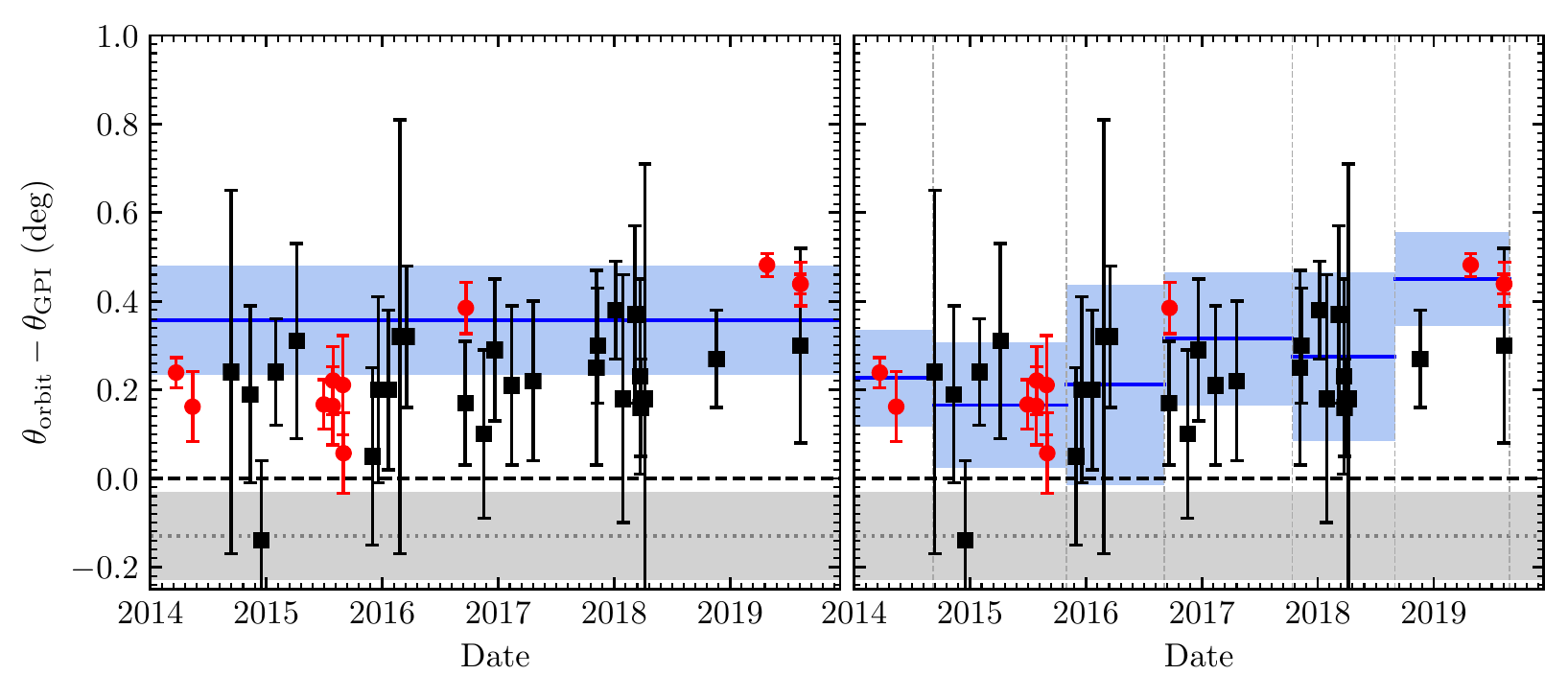}
\end{tabular}
\end{center}
\caption 
{ \label{fig:tn}
Measurements of the north offset angle of GPI derived from calibration binaries (red circles) and the $\theta^1$ Ori B2-B3 binary (black squares). We fit the north angle assuming it is either a constant calibration for the entire date range (left panel), or that it varies between telescope shutdowns (right panel). The mean and standard deviation (blue solid line and shaded region) are calculated as in Fig.~\ref{fig:ps}. The previous astrometric calibration is overplotted for reference (grey dashed line and shaded region).} 
\end{figure} 


The north offset angle for GPI was measured by taking the difference of the position angle of the companion predicted from the NIRC2-only orbit fit ($\theta_{\rm orbit}$) and the measured position angle within the reduced GPI data cubes ($\theta_{\rm GPI}$). This difference is reported in Table~\ref{tbl:tn_log} for each calibration binary measurement. We calculated a weighted mean of $0.36\pm0.12$\,deg for the full set of measurements, with the error calculated assuming that they were not independent. The aforementioned 0.12\,deg uncertainty includes a 0.1\,deg uncertainty that was added in quadrature to account for systematics and uncertainties in the relative astrometry. Measurements with large uncertainties in the predicted position angle ($\theta_{\rm orbit}$) were excluded. The measured offsets and the best fit model are plotted in Figure~\ref{fig:tn} (left panel). While the model is consistent with the measurements given the sizes of the uncertainties on both the measurements and the model ($\chi^2_{\nu} = 1.2$, $\nu=36$), there does appear to be a slight trend of increasing north offset angle over the course of six years when comparing the calibration binary measurements in early-2014 and mid-2019.

One plausible cause of a rotation of the instrument with respect to the telescope is the annual shutdown of the telescope when both the instrument and instrument support structure are removed to perform maintenance. We fit a variable north offset angle that remains static between the dates of telescope shutdowns. A series of weighted means were calculated using measurements between each shutdown, as listed in Table~\ref{tbl:tn_log} and plotted in Figure~\ref{fig:tn}. This model reproduces the trend of increasing north offset angle during the previous six years and is an improved fit ($\chi^2_{\nu} = 0.4$, $\nu=31$) relative to the single-valued model. We opted to use this variable north offset angle model for the final astrometric calibration of the instrument.

\subsection{Instrument Stability}
\begin{figure}
\begin{center}
\begin{tabular}{c}
\includegraphics[width=16cm]{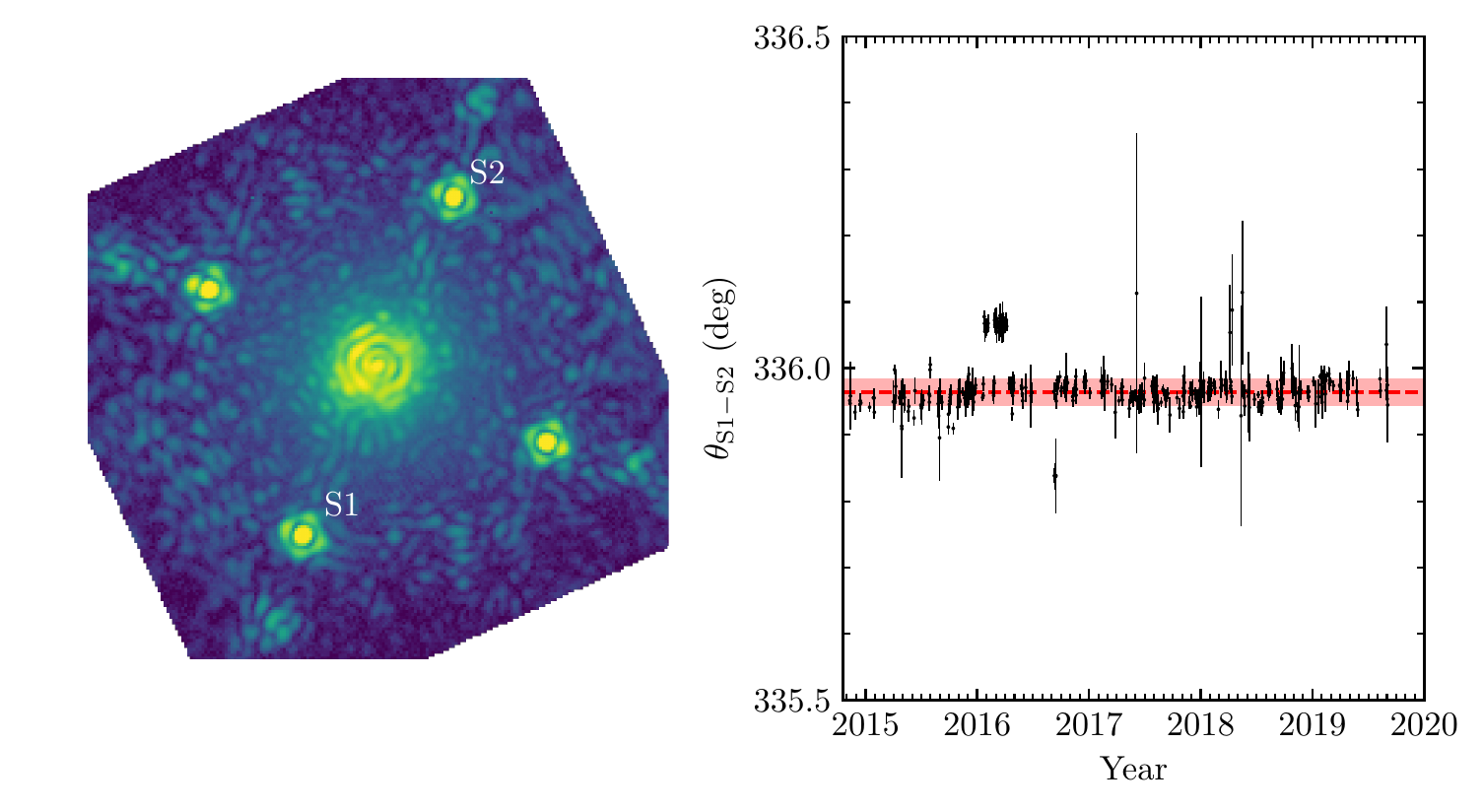}
\end{tabular}
\end{center}
\caption 
{ \label{fig:Hspot}
One wavelength slice of a reduced GPI data cube for a post-alignment image taken using GPI's internal source on 2014 November 12 (left panel). The four satellite spots generated by the grid on the pupil plane apodizer are clearly visible. The position angle between the bottom left (S1) and top right (S2) satellite spot, measured from S1 to S2 counter-clockwise from vertical, plotted as a function of date for each post-alignment image taken since the instrument was commissioned (right panel).} 
\end{figure}

The cause of the change of the north offset angle over time is not known. In principle, a movement of the IFS or the CAL system on their bipod mounts could produce a clocking of the focal plane with respect to the telescope, although a movement of ~5 mm would be required.  We excluded rotations internal to the instrument by measuring the angle between two of the satellite spots within a post-alignment image taken routinely before instrument operation. These satellite spots are generated by a periodic wire grid on the pupil plane apodizer\cite{Marois:2006kh,Sivaramakrishnan:2006cf}, located on the AO bench (Fig.~\ref{fig:gpi_diagram}). A physical rotation of the IFS relative to the apodizer would manifest itself as a rotation of the satellite spots within the focal plane as recorded by the IFS. We measured the angle between the bottom left and top right satellite spots in 406 post-alignment images taken between late-2014 and mid-2019 using the satellite spot finding algorithm that is a part of the GPI DRP. We find no significant trend in this angle over the past five years (Fig.~\ref{fig:Hspot}), although a significant offset of $\sim$0.1\,deg is seen for a few months at the start of 2016 that coincides with mechanical difficulties with the wheel containing the pupil plane apodizers. Excluding this period, we find an angle between these two satellite spots of $335.96\pm0.02$\,deg. The stability of this angle implies that the change in the north offset angle seen in Figure~\ref{fig:tn} is caused by a mechanical rotation upstream of the pupil plane mechanism containing the apodizer. The GPI optics upstream of this are all rigidly mounted in a single plane onto a thick optical bench and are extremely unlikely to produce such a rotation. In principal a rotation of the outer truss structure holding all three assemblies with respect to the mounting plate could rotate the focal plane, but again that would have to be on the order of 5 mm, essentially impossible. GPI has an extremely rigid truss structure supporting various subcomponents. Integrated FEA/optical modeling shows that flexure motions of any component relative to the optical axis are $<25$\,$\mu$m over the operating range of gravity vectors \cite{Erickson:2008hp}. Although we did not explicitly model rotation, if any hypothetical rotation component involves displacements on the same scale, the angular rotation would be on the order of 0.01\,deg. The pins that locate GPI onto the ISS face have much more precise tolerances than that as well ($<0.23$\,mm). 

\section{Revised Astrometry for Substellar Companions}
\label{sec:companion_astro}
\begin{figure}
\begin{center}
\begin{tabular}{c}
\includegraphics[width=16cm]{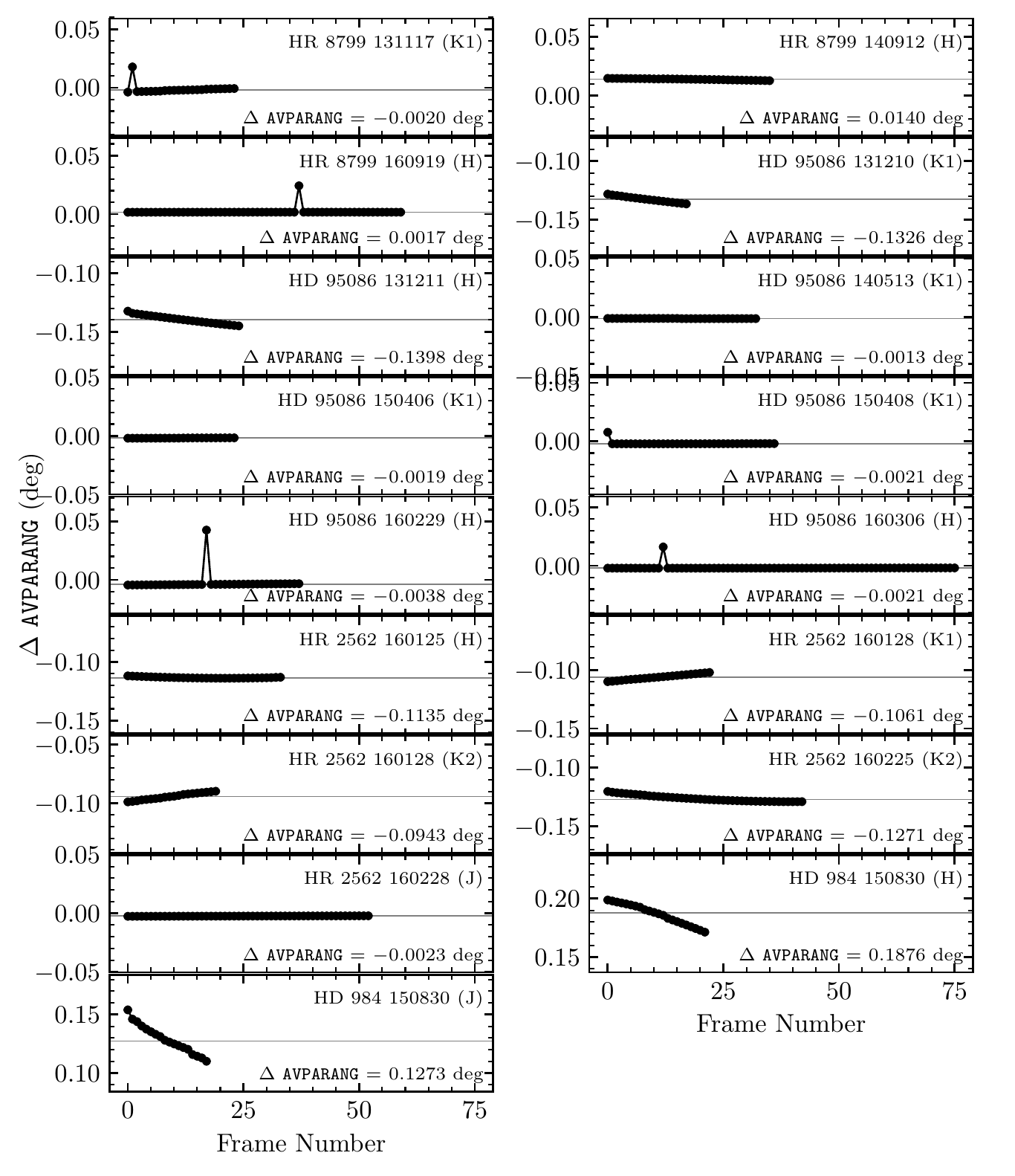}
\end{tabular}
\end{center}
\caption 
{ \label{fig:delta_pa} The change in the average parallactic angle header keyword ({\tt AVPARANG}) due to the revisions to the pipeline described in Sections~\ref{sec:updates}, \ref{sec:clocks}, and \ref{sec:rotator} for four GPIES targets that have published astrometry of substellar companions. The change in parallactic angle varies for each epoch, and for HD 984 varies significantly within a single epoch. The error in the integration described in Sec.~\ref{sec:romberg} is apparent in several epochs, most noticeably for the 2016 February 29 dataset on HD 95086 (middle row, left column).} 
\end{figure}

The changes to the pipeline described in Sections~\ref{sec:updates}, \ref{sec:clocks}, and \ref{sec:rotator} and the revised astrometric calibration of the instrument described in Section~\ref{sec:new_tn} both necessitate a revision of previously-published relative astrometry of substellar companions measured using GPI observations. Revisions for $\beta$~Pictoris~b \cite{Nielsen:td}, 51~Eridani~b \cite{DeRosa:2020gy}, and HD~206893~B (Ward-Duong et al. 2019, \textit{submitted}) are presented in other works. Here, we present corrections to the astrometry for the exoplanets in the HR 8799\cite{Wang:2018fd} and HD 95086\cite{Rameau:2016dx} systems, and the brown dwarfs HR 2562 B\cite{Konopacky:2016dk} and HD 984 B\cite{JohnsonGroh:2017kh}, that correct for the changes to the pipeline and the revised astrometric calibration of the instrument. We reduced the same images used in the previous studies with the latest version of the GPI DRP. The revisions described in Sections~\ref{sec:updates}, \ref{sec:clocks}, and \ref{sec:rotator} all affect the {\tt AVPARANG} header keyword. The change in this value is plotted as a function of frame number for each observing sequence in Figure~\ref{fig:delta_pa}. $\Delta$ {\tt AVPARANG} is typically small and static, only changing by at most $\sim0.05$\,deg between the start and end of the J-band sequence on HD 984 taken on 2015 August 30. The effect of the parallactic angle integration error described in Sec.~\ref{sec:romberg} is apparent in several epochs.

The median $\Delta$ {\tt AVPARANG} was used in conjunction with the revised north offset angle described in Sec.~\ref{sec:new_tn} to revise the previously-published astrometry. We assumed that a single offset to the measured position angle of a companion accurately describes the effect of the change to the parallactic angle for each frame within a sequence. As the maximum change in $\Delta$ {\tt AVPARANG} over a sequence was 0.05\,deg, the effect on the companion astrometry is likely on this order, or smaller. For the majority of cases $\Delta$ {\tt AVPARANG} changes by less than one one-hundredth of a degree over the course of a full observing sequence. The previous and revised astrometry for each published epoch are given in Table~\ref{tbl:revised_astrometry}. We find small but not significant changes in the measured separations, and significant changes in the measured position angles due to the significant change in the north offset angle described in Sec.~\ref{sec:new_tn}.

\begin{table}[ht]
\caption{Revised companion astrometry} 
\label{tbl:revised_astrometry}
\begin{center}       
\begin{tabular}{ccccccc}
Object & Date & Band & $\rho_{\rm original}$ & $\theta_{\rm original}$ & $\rho_{\rm revised}$ & $\theta_{\rm revised}$\\
&(UT)&&(mas)&(deg)&(mas)&(deg)\\
\hline
HR 8799 c & 2013-11-17 & K1 & $949.5\pm0.5$ & $325.18\pm0.14$ & $949.1\pm1.4$ & $325.51\pm0.12$ \\
HR 8799 d & 2013-11-17 & K1 & $654.6\pm0.9$ & $214.15\pm0.15$ & $654.3\pm1.3$ & $214.48\pm0.13$ \\
HR 8799 e & 2013-11-17 & K1 & $382.6\pm2.1$ & $265.13\pm0.24$ & $382.4\pm2.2$ & $265.46\pm0.23$ \\
HR 8799 b & 2014-09-12 & H & $1721.2\pm1.4$ & $65.46\pm0.14$ & $1720.5\pm2.8$ & $65.74\pm0.15$ \\
HR 8799 c & 2014-09-12 & H & $949.0\pm1.1$ & $326.53\pm0.14$ & $948.6\pm1.7$ & $326.81\pm0.15$ \\
HR 8799 d & 2014-09-12 & H & $662.5\pm1.3$ & $216.57\pm0.17$ & $662.2\pm1.6$ & $216.85\pm0.18$ \\
HR 8799 c & 2016-09-19 & H & $944.2\pm1.0$ & $330.01\pm0.14$ & $943.8\pm1.7$ & $330.43\pm0.16$ \\
HR 8799 d & 2016-09-19 & H & $674.5\pm1.0$ & $221.81\pm0.15$ & $674.2\pm1.4$ & $222.23\pm0.17$ \\
HR 8799 e & 2016-09-19 & H & $384.8\pm1.7$ & $281.68\pm0.25$ & $384.6\pm1.8$ & $282.10\pm0.26$ \\
HD 95086 b & 2013-12-10 & K1 & $619.0\pm5.0$ & $150.90\pm0.50$ & $618.9\pm4.9$ & $151.10\pm0.44$ \\
HD 95086 b & 2013-12-11 & H & $618.0\pm11.0$ & $150.30\pm1.10$ & $617.8\pm11.1$ & $150.45\pm1.11$ \\
HD 95086 b & 2014-05-13 & K1 & $618.0\pm8.0$ & $150.20\pm0.70$ & $617.7\pm8.0$ & $150.55\pm0.71$ \\
HD 95086 b & 2015-04-06 & K1 & $622.0\pm7.0$ & $148.80\pm0.60$ & $621.9\pm7.3$ & $149.06\pm0.64$ \\
HD 95086 b & 2015-04-08 & K1 & $622.0\pm4.0$ & $149.00\pm0.40$ & $621.7\pm4.1$ & $149.25\pm0.39$ \\
HD 95086 b & 2016-02-29 & H & $621.0\pm5.0$ & $147.80\pm0.50$ & $620.3\pm4.8$ & $148.09\pm0.57$ \\
HD 95086 b & 2016-03-06 & H & $620.0\pm5.0$ & $147.20\pm0.50$ & $619.8\pm4.8$ & $147.50\pm0.57$ \\
HR 2562 B & 2016-01-25 & H & $619.0\pm3.0$ & $297.56\pm0.35$ & $618.8\pm3.0$ & $297.76\pm0.40$ \\
HR 2562 B & 2016-01-28 & K1 & $618.0\pm5.0$ & $297.40\pm0.25$ & $617.8\pm5.1$ & $297.50\pm0.30$ \\
HR 2562 B & 2016-01-28 & K2 & $618.0\pm4.0$ & $297.76\pm0.37$ & $618.0\pm4.1$ & $297.88\pm0.42$ \\
HR 2562 B & 2016-02-25 & K2 & $619.0\pm2.0$ & $297.50\pm0.25$ & $618.9\pm2.1$ & $297.58\pm0.31$ \\
HR 2562 B & 2016-02-28 & J & $620.0\pm3.0$ & $297.90\pm0.25$ & $620.2\pm3.0$ & $298.11\pm0.32$ \\
HD 984 B & 2015-08-30 & H & $216.3\pm1.0$ & $83.30\pm0.30$ & $216.2\pm1.0$ & $83.76\pm0.30$ \\
HD 984 B & 2015-08-30 & J & $217.9\pm0.7$ & $83.60\pm0.20$ & $217.8\pm0.8$ & $84.00\pm0.21$ \\
\end{tabular}
\end{center}
\end{table} 

\section{Discussion/Conclusion}
We have identified and corrected several issues with the Gemini Planet Imager Data Reduction Pipeline that affected astrometric measurements of both calibration binaries and substellar objects whose orbital motion was being monitored. We reprocessed the calibration data after implementing these fixes into the pipeline, and revised the astrometric calibration of the instrument. The most significant change was to the north offset angle; changing from $-0.10\pm0.13$\,deg to between $0.17\pm0.14$\,deg and $0.45\pm0.11$\,deg, depending on the date. The plate scale of the instrument was also re-measured as $14.161\pm0.021$\,mas\,px$^{-1}$, consistent with the previous calibration albeit with a larger uncertainty.

While the change to the astrometric calibration of the instrument is significant relative to the stated uncertainties, the impact should be limited to studies that combine GPI astrometry with that from instruments of similar precision. The revised calibration should not have a significant impact on the results and interpretation of studies that used GPI astrometry either solely, or in conjunction with astrometry from instruments with significantly worse astrometric precision\cite{MillarBlanchaer:2015ha,Wang:2016gl,DeRosa:2015jl}; an offset in the north angle will simply change the position angle of the orbit on the sky ($\Omega$). A more significant effect might be seen for orbit fits that combined astrometry from GPI with astrometry of a similar precision from other instruments\cite{Rameau:2016dx,Maire:2019kb}. The magnitude of the effect on the derived orbital parameters is likely small. All but one of the substellar companions studied with GPI have a small fraction of their complete orbits measured, and so the change of the shape of the posterior distributions describing the orbital elements is likely not statistically significant. The precision of astrometric measurements made with GPI is currently limited by measurement uncertainties except for widely-separated companions such as the HR 8799 bcd, and the highest SNR measurements of $\beta$~Pic~b made in 2013 when the projected separation was $\sim430$\,mas, where the north angle uncertainty dominates the position angle error budget. Lower SNR measurements of faint companions such as 51 Eri b are less affected, with the north angle uncertainty being between a factor of two and five smaller than the measurement uncertainty.

Future studies using archival GPI data will need to account for both the changes to the pipeline and the revision to the astrometric calibration. The updated pipeline is publicly available on the Gemini Planet Imager instrument website\footnote{\url{ http://docs.planetimager.org/pipeline/ }} and on GitHub\footnote{\url{https://github.com/geminiplanetimager/gpi\_pipeline/ }}. All users wishing to perform precision astrometry will have to reduce their data using the latest version of the pipeline, especially those obtained on the highlighted dates in Figure~\ref{fig:ut_offset}, and apply the revised astrometric calibration presented in Sec.~\ref{sec:new_tn}. The measurements presented here demonstrate the importance of continued astrometric calibration, especially for instruments on the Cassegrain mount of a telescope. Improvements to the limiting magnitude of GPI's adaptive optics system as it is moved to Gemini North will allow us to use globular clusters as astrometric calibrations instead of isolated binaries, allowing for a more precise determination of the north angle via a comparison to both archival {\it Hubble Space Telescope} and contemporaneous Keck/NIRC2 observations.

This study also demonstrates the importance of precise and accurate astrometric calibration of instruments designed for high-contrast imaging of extrasolar planets. Instruments equipped with integral field spectrograph necessarily have a small field of view, challenging for astrometric calibration that typically relies on images of globular clusters extending over several to tens of arcseconds. These results also demonstrate the importance of accounting for orbital motion, either between the two components of a calibration binary, and/or the photocenter motion of one of the components if one of the components is itself a tight binary. A similar problem arises with the use of SiO masers near the Galactic Center\cite{Yelda:2010ig}; the location of the infrared source is not necessarily coincident with that of the radio emission that the infrared astrometric reference frame is tied to\cite{2019ApJ...873...65S}. Precise and accurate astrometric calibration of future instruments with very narrow fields of view such as the Coronagraphic Instrument (CGI) on the {\it Wide Field Infrared Survey Telescope}\cite{2015arXiv150303757S} will require a careful calibration strategy to mitigate the effects of these and other biases.

\subsection*{Disclosures}
The authors have no relevant financial interests and no other potential conflicts of interest to disclose.

\acknowledgments 
The authors wish to thank Brian Chinn, Carlos Quiroz, Ignacio Arriagada, Thomas Hayward, and Carlos Alvarez for useful discussions relating to this work. Supported by NSF grants AST-1411868 (R.D.R., E.L.N., K.B.F., B.M., and J.P.), AST-141378 (G.D.), AST-1518332 (R.D.R., J.J.W., T.M.E., J.R.G., P.G.K.), and AST1411868 (J.H., J.P.). Supported by NASA grants NNX14AJ80G (R.D.R., E.L.N., B.M., F.M., and M.P.), NSSC17K0535 (R.D.R, E.L.N., B.M., J.B.R.), NNX15AC89G and NNX15AD95G (R.D.R., B.M., J.E.W., T.M.E., G.D., J.R.G., P.G.K.). This work was performed under the auspices of the U.S. Department of Energy by Lawrence Livermore National Laboratory under Contract DE-AC52-07NA27344. This work benefited from NASA's Nexus for Exoplanet System Science (NExSS) research coordination network sponsored by NASA's Science Mission Directorate. J.R. is supported by the French National Research Agency in the framework of the Investissements d'Avenir program (ANR-15-IDEX-02). Based on observations obtained at the Gemini Observatory, which is operated by the Association of Universities for Research in Astronomy, Inc., under a cooperative agreement with the NSF on behalf of the Gemini partnership: the National Science Foundation (United States), National Research Council (Canada), CONICYT (Chile), Ministerio de Ciencia, Tecnolog\'{i}a e Innovaci\'{o}n Productiva (Argentina), Minist\'{e}rio da Ci\^{e}ncia, Tecnologia e Inova\c{c}\~{a}o (Brazil), and Korea Astronomy and Space Science Institute (Republic of Korea). Some of the data presented herein were obtained at the W. M. Keck Observatory, which is operated as a scientific partnership among the California Institute of Technology, the University of California and the National Aeronautics and Space Administration. The Observatory was made possible by the generous financial support of the W. M. Keck Foundation. The authors wish to recognize and acknowledge the very significant cultural role and reverence that the summit of Maunakea has always had within the indigenous Hawaiian community.  We are most fortunate to have the opportunity to conduct observations from this mountain. This work has made use of data from the European Space Agency (ESA) mission {\it Gaia} (\url{https://www.cosmos.esa.int/gaia}), processed by the {\it Gaia} Data Processing and Analysis Consortium (DPAC, \url{https://www.cosmos.esa.int/web/gaia/dpac/consortium}). Funding for the DPAC has been provided by national institutions, in particular the institutions participating in the {\it Gaia} Multilateral Agreement. This research has made use of the SIMBAD database and the VizieR catalog access tool, both operated at the CDS, Strasbourg, France. This research has made use of the Washington Double Star Catalog maintained at the U.S. Naval Observatory.


\bibliographystyle{spiejour}   

\listoffigures
\listoftables

\end{spacing}
\end{document}